\pdfoutput=1
\documentclass[11pt]{article}

\usepackage[preprint]{acl}
\usepackage{times}
\usepackage{latexsym}
\usepackage{booktabs}
\usepackage{algorithm}
\usepackage{algpseudocode}
\usepackage{amsmath}
\usepackage{multirow}
\usepackage{makecell}
\usepackage[T1]{fontenc}
\usepackage[utf8]{inputenc}

\usepackage{microtype}

\usepackage{inconsolata}

\usepackage{graphicx}
\usepackage{amssymb}

\definecolor{blueyuhan}{RGB}{0,0,255}

\title{\textit{Thinking Before Running}! Efficient Code Generation with \\ Thorough Exploration and Optimal Refinement}

\author{
\begin{tabular}{c}
Xiaoqing Zhang$^{1,2}$\thanks{This work was done during the internship at Moonshot AI.} \quad \quad 
Yuhan\  Liu$^{1}$\footnotemark[2] \quad \quad  \textbf{Flood Sung}$^{2}$ \\\quad \textbf{Xiuying Chen}$^{3}$ \quad \quad \ \ \textbf{Shuo \ Shang}$^{4}$ \quad \quad \textbf{Rui \ Yan}$^{1,5,6}$\thanks{\ \ Corresponding authors.} 
\end{tabular}
\\ \vspace{.5mm}
    \small
    \begin{tabular}{c}
    $^1$Gaoling School of Artificial Intelligence, Renmin University of China \quad $^2$Moonshot AI\\ $^3$Mohamed bin Zayed University of Artificial Intelligence\\
    $^4$University of Electronic Science and Technology of China \quad $^5$School of Artifcial Intelligence, Wuhan University\\
    $^6$Engineering Research Center of Next-Generation Intelligent Search and Recommendation, MoE\\
    \end{tabular}
    \\ \vspace{.5mm}
    \small
    \begin{tabular}{c}
    \texttt{\{xiaoqingz, yuhan.liu, ruiyan\}@ruc.edu.cn} \quad \texttt{floodsung@moonshot.cn}\\
    \texttt{xy-chen@pku.edu.cn}\quad \texttt{jedi.shang@gmail.com} \\
    \end{tabular}
    \vspace{2mm} \\
}

\begin{document}
\maketitle
\begin{abstract}
Code generation is crucial in software engineering for automating the coding process efficiently. 
While test-time computation methods show promise, they suffer from high latency due to multiple computation rounds.
To overcome this, we introduce \textbf{ThinkCoder}, a framework that combines thorough exploration with optimal refinement.
The exploration phase diversifies the solution space by searching for potential solutions, followed by a refinement phase that enhances precision.
This approach allows us to select the best solution through careful consideration before taking action, avoiding excessive trial and error.
To further minimize test-time computation overhead, we introduce preference-driven optimization with Reinforced Self-Training (ReST), which uses exploration trajectories from ThinkCoder to guide LLM's evolution.
This approach enhances LLM's exploration efficiency via preference learning, cutting costs while maintaining accuracy.
ThinkCoder boosts the performance with a single LLM, excelling on benchmarks like HumanEval and MBPP. 
Compared to SOTA models, it improves Pass@1 by 3.0\% over MapCoder with just 6.4\% of the computation cost.
Against AgentCoder, ThinkCoder achieves a 0.5\% higher Pass@1 after 2 rounds, outperforming AgentCoder's 5 rounds.
Additionally, ReST with success trajectories enhances efficiency, allowing models like LLaMA2-7B to achieve competitive results using only 20\% of the computational resources. 
These results highlight the framework's effectiveness and scalability. 
\footnote{The code is available at \url{https://github.com/xiaoqzhwhu/ThinkCoder}}
\end{abstract}

\section{Introduction}

Recent research advances indicate that large language models (LLMs) have demonstrated remarkable capabilities in various programming-related domains, such as code generation~\cite{zheng2023survey,chaudhary2023code,dong2024self,li2023structured}, code refinement~\cite{chen2023improving,guo2024exploring,zheng2024opencodeinterpreter,ridnik2024code,liu2024refining,zhang2024pair}, and software testing~\cite{li2023finding,jalil2023chatgpt,wang2024software,jones2024automatic}. 
By utilizing extensive training data, LLMs can understand complex programming tasks, generate syntactically accurate code, and even enhance the quality of their solutions through iterative improvement.

%drawback 2: integrate feedback needs plenty of tests 
%drawback 1: multi-rounds
% In the context of code generation, researchers have made substantial progress in optimizing prompt design, test case optimization, and feedback integration. 
% These techniques are vital for enhancing the precision and diversity of generated code. 
% However, achieving high-quality outputs is still a significant challenge, as many current approaches depend heavily on large-scale models with extensive instruction-following capabilities.
In the context of code generation, researchers have made significant progress in improving prompts to encourage LLMs to generate higher-quality answers. This includes two main directions: optimizing the path for answer exploration and incorporating rich reflection information.
In the first direction of path optimization, for example, Self-Planning~\cite{jiang2024self} reduces problem complexity in code generation by introducing a planning phase, which refines the solution generation process. Similarly, CodeCoT~\cite{huang2023codecot} improves the exploration by employing a Chain-of-Thought(CoT) prompt, enabling models to reason through each step systematically. 
In the second direction, the integration of reflection, test cases, and feedback is progressively being used to steer and enhance the generation process.
For instance, RethinkMCTS~\cite{li2024rethinkmcts} integrates verbal feedback into its Monte Carlo Tree Search framework to correct errors and improve code generation. Similarly, AgentCoder~\cite{huang2023agentcoder} boosts efficiency and accuracy by using interactions between multiple agents to provide valuable input.
% —Programmer, Test Designer, and Test Executor—to provide valuable input.
% This framework also heavily depends on feedback from test cases for code correction, and inaccuracies in feedback or limited iteration cycles can result in erroneous final outputs.
% However, these systems are similarly limited by the quality and quantity of feedback and the number of iterative steps they can afford to take.

\textcolor{black}{Despite the significant potential of using CoT for planning code improvements, the thought process incurs a substantial token overhead. 
To enhance efficiency and minimize trial-and-error costs, it is crucial to clarify the optimization direction before implementing refinement. 
Existing methods, such as MapCoder, which leverages four LLM-based agents, consume a considerable amount of computational resources.}

Based on these insights, we propose ThinkCoder, a novel framework for efficient code generation that combines thorough exploration and optimal refinement. 
First, we simplified the multi-agent framework into a single Exploration Agent and a \textcolor{black}{CodeVerifier}. 
\textcolor{black}{The Exploration Agent is responsible for in-depth analysis and generating various codes, while exploring diverse test cases to build a robust testing pool. 
The CodeVerifier, using the testing pool, independently determines the optimization direction without relying on LLMs. By continuously improving the testing pool throughout the refinement process, ThinkCoder ensures accurate verification of generated code in each round, providing efficient guidance for subsequent optimizations.}
% Despite their impressive capabilities, a critical challenge remains in fully exploiting the potential of LLMs for high-quality code generation, as their performance is still constrained by issues such as high computation costs and limited effectiveness in integrating iterative feedback. 
% To enhance efficiency and minimize trial-and-error costs, it's essential to define the improvement direction clearly before implementing refinement. While existing methods like MapCoder, which leverages four LLM-based agents, consumes a considerable amount of computational resources.

% To address these challenges, we propose ThinkCoder, a novel framework for efficient code generation that combines thorough exploration and optimal refinement.
% First, we simplified the multi-agent framework into a single Exploration Agent and a \textcolor{black}{CodeVerifier}. 
% The Exploration Agent is responsible for deep thinking and expanding exploration with various codes and tests.
% The CodeVerifier independently verifies code during the optimal refinement process, without relying on LLMs. 
% By utilizing a dynamically created testing pool, ThinkCoder ensures accurate verification of the generated code and offers reliable guidance for further optimization. 
Second, to tackle the additional computational load resulting from extensive exploration, we incorporate a preference-driven optimization phase using Reinforced Self-Training (ReST) in ThinkCoder. 
ReST leverages successful exploration trajectories to train LLM, enabling it to generate optimal solutions more efficiently during exploration. 
This approach significantly reduces the computational cost of test-time computations, enabling models like LLaMA2-7B to achieve notable improvements in both performance and efficiency.
% we propose a test-time compute-based framework, RPSVR. 

% Figure \ref{tab:intro} illustrates the difference in adopted techniques between RPSVR and other methods. 
% Even when the model cannot accurately fix code based on test feedback, it can improve the likelihood of reaching correct solutions through diverse code exploration. We introduce a self-generated testing pool, which serves two purposes: verifying the correctness of diverse solutions and enhancing feedback quality through more comprehensive test cases, including functional tests, performance tests, and boundary tests. Our RPSVR framework leverages repeated exploration and validation to progressively enhance code generation quality. The test-time compute method improves code quality but also increases inference time. To mitigate this cost, we collected beneficial exploration trajectories and applied ReST training to LLaMA2-7B, enabling the model to learn how to generate optimal solutions faster during exploration while providing higher-quality test cases. This approach shortens the exploration cycle and enhances overall efficiency.

Our contributions are as follows:

$\bullet$ \textcolor{black}{We introduce ThinkCoder, an efficiency framework for code generation that integrates an LLM-based exploration agent with a non-LLM-based CodeVerifier. We utilize a self-evolving testing pool, significantly enhancing the speed of code verification and improvement.}

$\bullet$ We introduce preference-driven optimization with ReST, using exploration trajectories from ThinkCoder to enhance LLMs' generation and verification capabilities, further reducing computational overhead during testing.

$\bullet$ We have demonstrated the effectiveness of ThinkCoder on multiple benchmark datasets and across different scales of LLMs, achieving state-of-the-art performance in code generation while optimizing computational efficiency.
% across multiple code benchmarks and LLMs of various scales. We have achieved state-of-the-art performance in code generation by optimizing the test-time compute budget.

\section{Related Work}

\textbf{Test-time Compute.}
Test-time computing methods use self-play and additional verifiers to enhance performance, despite the computational overhead. For example, training a Verifier with Preference Reward Model data and combining strategies like Best-of-N Weighted, Beam Search, and Lookahead Search can help \cite{snell2024scaling}. ToolLLM \cite{qin2023toolllm} evaluates tool usage by comparing success rates and solution quality to ChatGPT-ReACT, employing ToolEval as a Verifier for optimal API calls using depth-first search. However, API-based Verifiers are limited by training data and struggle with generalization.
In contrast, our approach employs a task-based testing tool as a Verifier, offering a simpler and more effective solution that is closely aligned with the specific task requirements.

\textbf{Multi-Agent Code Generation.}
Deriving multiple agents by mimicking the different stages of human programming can effectively improve both the quality and efficiency of code generation, as agents based on large language models possess strong reasoning capabilities~\cite{liu2024skepticism,liu2024tiny,zhang2025weaving}. For instance, AgentCoder~\cite{huang2023agentcoder} introduces programmer agents, test design agents, and test execution agents during the code generation process. Similarly, MapCoder~\cite{islam2024mapcoder} introduces retrieval agents, planning agents, coding agents, and debugging agents, iterating on code generation quality through more frequent agent communication and reflection. 
However, MapCoder incurs higher computational overhead and relies solely on sample I/O for code execution, limiting test case coverage and robustness. In contrast, our framework, ThinkCoder, addresses these limitations by self-verification efficiently without excessive computational costs.

\begin{figure*}[htb]
  % \begin{wrapfigure}{r}{0.48\textwidth}
  \centering
  \includegraphics[width=0.9\textwidth]{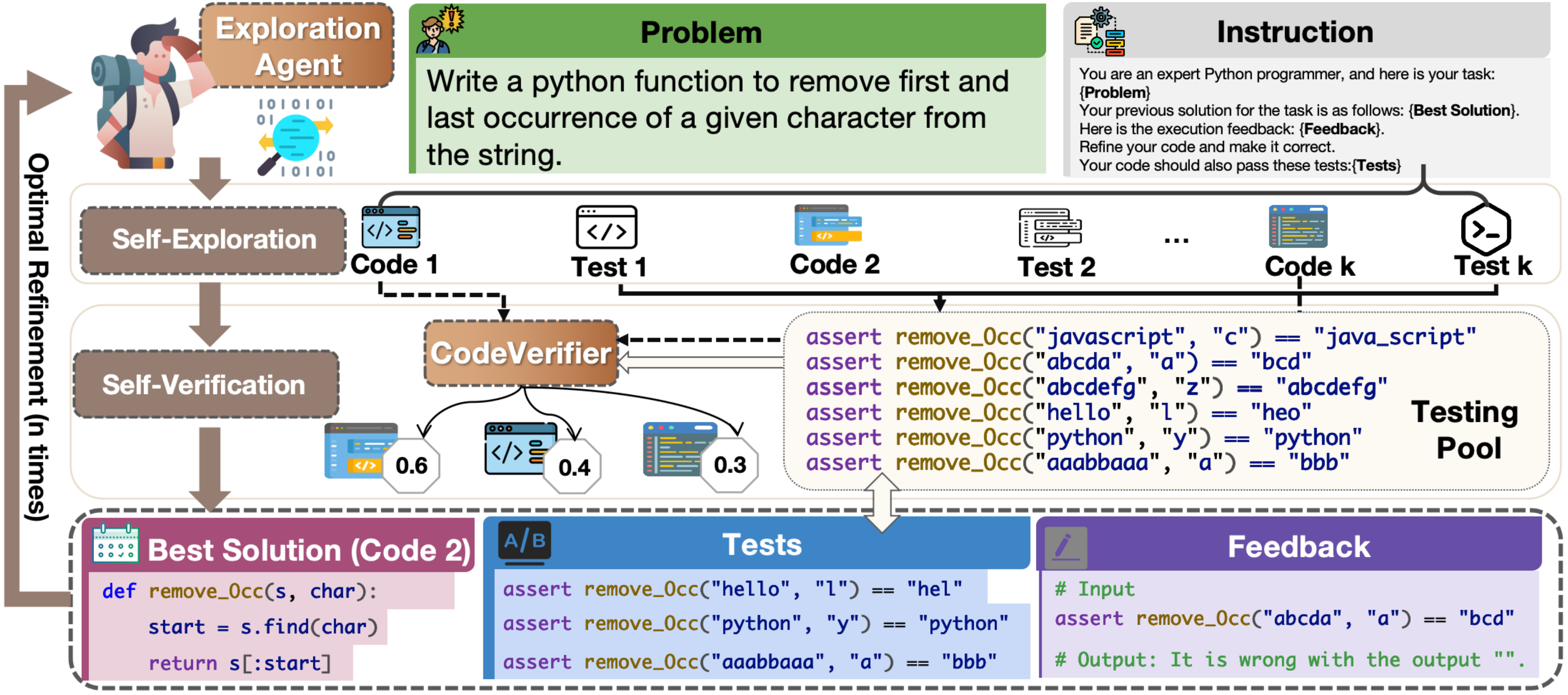}
  \caption{The end-to-end process of ThinkCoder involves $k$ thorough exploration steps followed by $n$ optimal refinement cycles. The Exploration Agent generates $k$ codes and tests simultaneously during self-exploration, storing results in a Testing Pool. The optimal refinement phase includes self-verification that selects the optimal solution with the CodeVerifier and aggregate reflection for the instruction update. The optimal refinement will be repeated recursively for $n$ cycles, ultimately leading to the final solution.}
  \label{fig:framework}
% \end{wrapfigure}
\end{figure*}

\textbf{Instruction Tuning for Code.}
High-quality code generation models often require instruction fine-tuning based on large-scale pre-trained language models. The fine-tuning data generally comes from real-world collections or is artificially synthesized. 
For example, Code Alpaca~\cite{chaudhary2023code} combines the SELF-INSTRUCT strategy with code fine-tuning, CODELLAMA~\cite{roziere2023code} generates questions, solutions, and test cases through prompts, and OctoPack~\cite{muennighoff2023octopack} directly collects data from Git. 
LLMs fine-tuned with instructions possess stronger code generation capabilities. 
% However, the schemes for collecting instruction data, the volume of data, and fine-tuning strategies lack detailed manuals. 
% We have published a detailed strategy for fine-tuning Llama2-7B based on instruction-tuning data collected using CodeQwen~\cite{bai2023qwen}, providing a completely transparent solution to support the advancement of this field.
However, existing methods often lack comprehensive guidelines for data collection, volume management, and fine-tuning strategies. To address this gap, we present a detailed fine-tuning strategy for Llama2-7B using instruction-tuning data collected via CodeQwen~\cite{bai2023qwen}, providing a transparent and reproducible methodology that supports the advancement of instruction-tuned code generation models.

\section{ThinkCoder}
\begin{figure*}[htb]
  % \begin{wrapfigure}{r}{0.48\textwidth}
  \centering
  \includegraphics[width=0.85\textwidth]{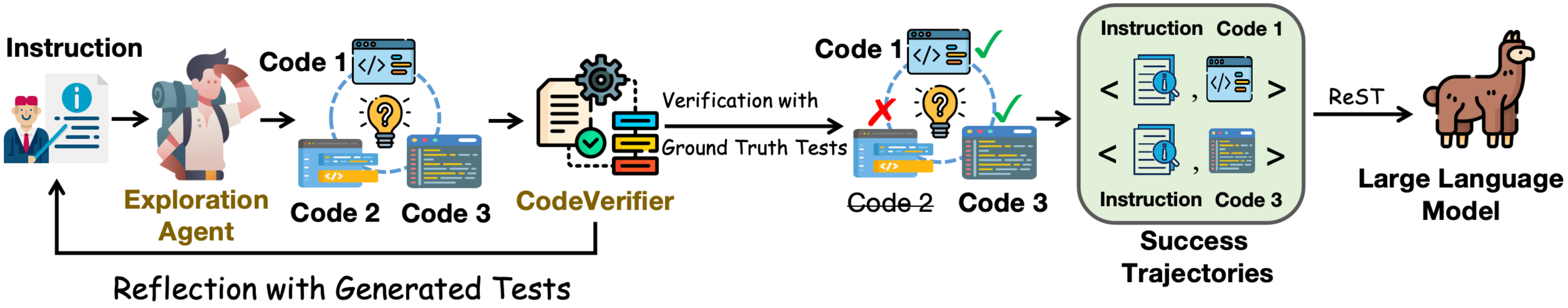}
  \caption{Trajectory collection with ThinkCoder and its application in ReST training for LLMs. We collect success trajectories offline based on the verification with ground truth tests, ensuring the solution aligns with human preferences. For reflection, we use LLM-generated tests, ensuring LLMs specifically address their own mistakes.}
  \label{fig:lora}
% \end{wrapfigure}
\end{figure*}

\subsection{Overall Architecture}
Figure \ref{fig:framework} illustrates the structure of ThinkCoder.
Given an instruction $I = [p, \text{prompt}]$ for problem $p$, where $\text{prompt}$ is the task description for code generation, the Exploration Agent produces $k$ candidate solutions. Each solution is paiblue with $m$ test cases, and the temperature $t$ of the LLM’s generation is adjusted to promote diversity. These test cases are then aggregated into a Testing Pool $TP$.
We define the exploration results as $E = [G, TP]$, where $G = \{ g_i \mid g_i = \text{LLM}(p, t) \text{ for } i \in \{1, 2, \dots, k\} \}$, and $TP \in \mathbb{T}^{k \times m}$, where $\mathbb{T}$ represents each test.
The CodeVerifier sequentially executes each candidate solution in $G$ using the test cases from $TP$ to compute pass rates. The code $g_s$ with the highest pass rate is selected for further refinement.
Here we define the pass rate of $g_i$ as $r_{g_i} = \frac{\text{number of tests passed by } g_i}{|TP|}$ and $g_s = \arg\max_{i} r_{g_i}$.
Feedback $f$ from failed test case $f_t$, along with the current code $g_s$, generated test cases $tests$, and problem $p$, is used to update the instructions $I=[p, g_s, f_t, f, \text{tests}, \text{prompt}]$ for the next exploration. 
This iterative process allows the Exploration Agent to generate improved solutions and test cases over $n$ iterations.
To avoid unnecessary computation on well-solved problems, if $r_{g_i}$ is greater than $\theta$, the refinement process is terminated, and the code $g_s$ is returned as the final solution.
For more details of ThinkCoder please refer to the algorithm \ref{alg:thinkcoder} in Appendix \ref{sec:algorithm}.
Below, we detail the implementation of Thorough Exploration, Optimal Refinement, Exploration Agent\&CodeVerifier, and Testing Pool.

\subsection{Thorough Exploration}
During the thorough exploration phase, ThinkCoder conducts extensive thinking before running code and performing corrections. The self-exploration process generates $k$ answers simultaneously to expand the search area by increasing the LLM's temperature, which flattens the probability distribution and enhances diversity. 
The first iteration of self-exploration aims to generate diverse code and comprehensive test cases, including regular tests, boundary tests, and performance tests. 
While the subsequent iterations, focus on the exploration of code correction for specific errors.
The only LLM-based agent conducts self-exploration with prompt design detailed in Appendix \ref{sec:prompts}.

\subsection{Optimal Refinement}
To solve the problem efficiently, it is essential to first identify the optimal solution through careful consideration before making any code corrections. 
Therefore, optimal refinement involves selecting the best solution via self-verification and gathering the necessary information for deeper exploration and improvement in the next exploration.
During self-verification, code quality is assessed by the pass rate, calculated by the CodeVerifier running candidate code against all test cases. The code with the highest pass rate is chosen for refinement. 
We then randomly select an unresolved error to target in the current round of improvements, ensuring all errors are eventually addressed.
The refinement instruction is generated based on the problem, the current best solution, feedback, and test cases. 
After each refinement, the solution’s pass rate is evaluated. If it exceeds previous solutions, it becomes the new best. 
Through multiple iterations, the best solution is continuously corrected and optimized, resulting in a gradual increase in the pass rate within the testing pool.
For more details, please refer to the Appendix \ref{sec:prompts}.

\subsection{Exploration Agent\& CodeVerifier}
ThinkAgent consists of a single exploration agent and a CodeVerifier. The exploration agent uses LLMs to search for diversity code and test cases, while the CodeVerifier runs the provided code snippets and test cases in a local environment to generate feedback.
To improve the efficiency of the CodeVerifier, we leverage multiprocessing to validate the code in parallel.

\subsection{Testing Pool}
The Testing Pool includes numerous test cases generated by the exploration agent, categorized into regular, boundary, and performance tests. 
\textcolor{black}{Our goal is to create a robust pool to assess complex situations beyond simple cases. 
A higher pass rate indicates stronger code robustness, so we ensure test case diversity and accuracy.} 
Several key operations contribute to achieving this. 
Abstract Syntax Tree (AST) ~\cite{backus1960report} is applied to identify and remove duplicate test cases in the testing pool, eliminating unnecessary time costs associated with redundant tests.
Additionally, after each iteration, the testing pool only selects test cases that can improve the code's pass rate, thus enhancing the diversity and accuracy of the pool. 
Finally, the testing pool will select test cases with distinct errors and use feedback to update the code's improvement goals, driving continuous optimization.

\section{Preference-driven Optimization}
In ThinkCoder, the generated code evolves as the pass rate on the testing pool improves. 
Each increase in the pass rate signifies the development of better codes. 
This progression demonstrates that the exploration agent is gradually producing higher-quality outputs. 
By collecting the trajectory of these improvements, we aim to fine-tune the model, enabling the LLM to generate superior code and test cases more effectively. 
This approach helps to refine the exploration process, clarify its direction, and reduce the time required for exploration.
Figure \ref{fig:lora} describes the process of trajectory collection and fine-tuning of the LLM.
We will explain this process from the perspectives of Data Collection, the Model Training and Inference.

\renewcommand{\arraystretch}{0.8}
\begin{table*}[htbp]
  \centering
  \resizebox{0.9 \textwidth}{!}{
\begin{tabular}{ccccccc}
\toprule
\textbf{Model} & \textbf{n} & \textbf{MBPP$\uparrow$} & \textbf{MBPP-ET$\uparrow$} & \textbf{HumanEval$\uparrow$} & \textbf{HumanEval-ET$\uparrow$} & \textbf{Avg$\uparrow$} \\
\midrule
\multirow{4}[8]{*}{\textbf{LLama2-7B-Chat}} & 0     & 27.8  & 22.9     & 12.9  & 11.2  & 18.7 \\
\cmidrule{2-7}      & 1     & 35.3(27.0\%)  & 29.0(26.6\%)     & 15.9(23.3\%)  & 11.6(3.6\%)  & 23.0(23.0\%) \\
\cmidrule{2-7}      & 2     & 37.5(34.9\%)  & 29.5(28.8\%)     & 17.1(32.6\%)  & 12.2(8.9\%)  & 24.1(28.9\%) \\
\cmidrule{2-7}      & 5     & \textbf{38.1(37.1\%)}  & \textbf{29.8(30.1\%)}     & \textbf{21.3(65.1\%)}  & \textbf{12.9(15.2\%)}  & \textbf{25.5(35.4\%)} \\
\midrule
\multirow{4}[8]{*}{\textbf{CodeQwen1.5-7B-Chat}} & 0     & 73.1  & 67.8     & 79.5  & 71.0    & 72.9 \\
\cmidrule{2-7}      & 1     & 80.2(9.7\%)  & 74.7(10.2\%)     & 90.9(14.3\%)  & 80.5(13.4\%)  & 81.6(11.9\%) \\
\cmidrule{2-7}      & 2     & 84.4(15.5\%)  & 75.0(10.6\%)     & 92.1(15.8\%)  & 80.5(13.4\%)  & 83.0(13.9\%) \\
\cmidrule{2-7}      & 5     & \textbf{86.0(17.6\%)}    & \textbf{75.0(10.6\%)}     & \textbf{93.3(17.4\%)}  & \textbf{80.5(13.4\%)}  & \textbf{83.7(14.8\%)} \\
\midrule
\multirow{4}[8]{*}{\textbf{Kimi}} & 0     & 70.0    & 67.4     & 81.6  & 73.4  & 73.1 \\
\cmidrule{2-7}      & 1     & 77.2(10.3\%)  & 73.5(9.1\%)     & 87.2(6.9\%)  & \textbf{79.9(8.9\%)}  & 79.5(8.8\%) \\
\cmidrule{2-7}      & 2     & 78.7(12.4\%)  & 74.5(10.5\%)     & \textbf{87.2(6.9\%)}  & 79.3(8.4\%)  & 79.9(9.3\%) \\
\cmidrule{2-7}      & 5     & \textbf{79.5(13.6\%)}  & \textbf{74.8(11.0\%)}     & 86.6(6.1\%)  & 79.3(8.4\%)  & \textbf{80.1(9.6\%)} \\
\midrule
\multirow{4}[8]{*}{\textbf{GPT-4-Turbo}} & 0     & 81.2  & 74.4     & 85.7  & 81.0  & 80.6 \\
\cmidrule{2-7}      & 1     & 89.5(5.1\%)  & 78.2(5.1\%)     & 90.2(5.3\%)  & 84.8(4.7\%)  & 85.7(6.3\%) \\
\cmidrule{2-7}      & 2     & 89.5(5.1\%)  & 79.9(7.4\%)     & 90.8(6.0\%)  & 84.8(4.7\%)  & 86.3(7.1\%) \\
\cmidrule{2-7}      & 5     & \textbf{89.5(5.1\%)}  & \textbf{80.0(7.5\%)}     & \textbf{91.5(6.8\%)}  & \textbf{84.8(4.7\%)}  & \textbf{86.5(7.3\%)} \\
\midrule
\multirow{4}[8]{*}{\textbf{GPT-4o}} & 0     & 84.9     & 78.2     & 92.5     & 85.8     & 85.4 \\
\cmidrule{2-7}      & 1     & 88.5(4.2\%)     & \textbf{79.9(2.2\%)}     & 96.3(4.1\%)     & 87.8(2.3\%)     & 88.1(3.2\%) \\
\cmidrule{2-7}      & 2     & 89.6(5.5\%)     & 79.2(1.3\%)     & 97.0(4.9\%)     & 89.0(3.7\%)     & 88.7(3.9\%) \\
\cmidrule{2-7}      & 5     & \textbf{90.0(6.0\%)}     & 79.2(1.3\%)     & \textbf{97.0(4.9\%)}     & \textbf{89.6(4.4\%)}     & \textbf{89.0(4.2\%)} \\
\bottomrule
\end{tabular}}%
    \caption{ThinkCoder performance of various LLMs. $n=0$ indicates results without the framework. $n>0$ indicates results after $n$ iterations of ThinkCoder, with each exploration generating $k=5$ solutions for optimal refinement at temperature $t=0.5$. Percentages in parentheses show improvement over the base LLMs.}
  \label{tab:rpsvr}%
  \vspace{-0.2cm}
\end{table*}%

\subsection{Data Collection}
The process of data collection for LLM fine-tuning involves two key steps. 
First, an exploration agent, denoted as $\mathcal{M}_0$, generates diverse low-probability data $D_{M_0}$ by sampling at various temperatures. 
Second, this data is further refined using reflection to produce an enhanced dataset $R_{M_0}$. 
In our data generation task, offline data collection starts with a training set $D = \{x_i\}_{i=1}^{N}$, where $N$ is the number of instances in the dataset. 
We generate high-quality preference data $D_{M_0} \cup R_{M_0}$ for LLM fine-tuning through the temperature-based and reflection-based generation process.

\textbf{Temperature-based Generation.}
During temperature-based generation, we obtain $D_{M_0}$ to enhance the diversity of responses.
For a given problem $x$, the language model $\mathcal{M}_0$ is used to explore $k$ times, generating $\{ y^j \}_{j=1}^k$, where $y^j \sim \mathcal{M}_0(y|x)$. 
During exploration, a relatively large temperature value $t$ is set for $\mathcal{M}_0$. 
The generated outputs are then evaluated using the CodeVerifier, which computes $E(x, y^j) = (\text{pass\_rate, feedback})$. 
For any sample where $\text{pass\_rate == 1}$, the corresponding instance $(x, y^j)$ is added to the training dataset $D_{M_0}$.

\textbf{Reflection-based Generation.}
The primary purpose of generating $R_{M_0}$ is to enhance the ability of error correction. 
For a given sample $x$ and the feedback $E(x, y^j) = (\text{pass\_rate, feedback})$ from the CodeVerifier, for samples with $\text{pass\_rate = 0}$, we add the feedback to $x$ and use the language model $\mathcal{M}_0$ to perform temperature-based exploration on the input $x' = (x, \text{feedback})$ and get the output $y'^j$.
Then we also use the CodeVerifier to obtain the right correction answer with $E(x', y'^j) = (\text{pass\_rate}', \text{feedback}')$. 
Finally, we add the instance $(x', y'^j)$ with the $\text{pass\_rate}' == 1$ to the reflection-based trajectories $R_{M_0}$.

\begin{table*}[htbp]
  \centering
  \scriptsize
  % \resizebox{12cm}{!}{
% Table generated by Excel2LaTeX from sheet 'all tables'
\resizebox{15cm}{!}{\begin{tabular}{ccccccc}
\toprule
& \textbf{Model} & \textbf{MBPP$\uparrow$}  & \textbf{MBPP-ET$\uparrow$} & \textbf{HumanEval$\uparrow$} & \textbf{HumanEval-ET$\uparrow$} & \textbf{Avg$\uparrow$} \\
% \textbf{Reflection} & 91    &  -     & 77.1  &  -     & 84.1 \\
% \midrule
% \textbf{ChatDev} & 84.1  &   -    & 79.8  &   -    & 82 \\
% \midrule
% Self-Collaboration & 90.2  & 70.7  & 78.9  & 62.1  & 75.5 \\
\midrule
\multirow{3}[2]{*}{\textbf{GPT-3.5-Turbo}} & MapCoder & 78.3  & 54.4  & 80.5  & 70.1  & 70.8 \\
% \midrule
& AgentCoder & 89.9  & 89.1    & 79.9  & 77.4  & 84.1 \\
% \midrule
& \textbf{ThinkCoder(Ours)} & \textbf{90.9}  &   78.8    & \textbf{89.0}    &   \textbf{81.0}      & \textbf{84.9} \\
\midrule
\multirow{3}[2]{*}{\textbf{GPT-4-Turbo}} & MapCoder & 83.1  & 57.7  & 93.9  & 82.9  & 79.4 \\
% \midrule
& AgentCoder & 91.4  & 91.4    & 89.6  & 76.2  & 87.2 \\
% \midrule
& \textbf{ThinkCoder(Ours)} & 90.5  &   80.5    & \textbf{93.9}    &   \textbf{86.0}      & \textbf{87.7} \\
\bottomrule
\end{tabular}}%
    % }
    \caption{Pass@1 for MapCoder, AgentCoder, and ThinkCoder on MBPP, MBPP-ET, HumanEval, and HumanEval-ET benchmarks with GPT-3.5-Turbo and GPT-4-Turbo as the backbone.}
  \label{tab:sota1}%
  % \vspace{-0.2cm}
\end{table*}%

\begin{table*}[htbp]
  \centering
  \scriptsize
% Table generated by Excel2LaTeX from sheet 'all tables'
\resizebox{15cm}{!}{\begin{tabular}{cccccccc}
\toprule
      & \textcolor{black}{\textbf{Model}} & \textcolor{black}{\textbf{MBPP$\uparrow$}} & \textcolor{black}{\textbf{MBPP-ET$\uparrow$}} & \textcolor{black}{\textbf{HumanEval$\uparrow$}} & \textcolor{black}{\textbf{HumanEval-ET$\uparrow$}} & \textcolor{black}{\textbf{CodeContests$\uparrow$}} & \textcolor{black}{\textbf{Avg$\uparrow$}} \\
\midrule
\multirow{3}[2]{*}{\textcolor{black}{\textbf{GPT-4o}}} & \textcolor{black}{LDB}   & \textcolor{black}{82.4}  & \textcolor{black}{65.4}  & \textcolor{black}{98.2}  & \textcolor{black}{81.7}  & \textcolor{black}{29.3}  & \textcolor{black}{71.4} \\
      & \textcolor{black}{LPW}   & \textcolor{black}{84.8}  & \textcolor{black}{65.8}  & \textcolor{black}{98.2}  & \textcolor{black}{84.8}  & \textcolor{black}{34.7}  & \textcolor{black}{73.7} \\
      & \textcolor{black}{\textbf{ThinkCoder(Ours)}} & \textcolor{black}{\textbf{90.0}} & \textcolor{black}{\textbf{79.2}} & \textcolor{black}{97.0}    & \textcolor{black}{\textbf{89.6}} & \textcolor{black}{\textbf{40.1}} & \textcolor{black}{\textbf{79.2} }\\
\bottomrule
\end{tabular}}%
    % }
    \caption{\textcolor{black}{Pass@1 for LDB, LPW, and ThinkCoder on MBPP, MBPP-ET, HumanEval, HumanEval-ET, and CodeContests benchmarks with GPT-4o as the backbone.}}
  \label{tab:sota2}%
  % \vspace{-0.2cm}
\end{table*}%

% \begin{table*}[htbp]
%   \centering
%   \scriptsize
% % Table generated by Excel2LaTeX from sheet 'all tables'
% \resizebox{15cm}{!}{\begin{tabular}{cccccccc}
% \toprule
%       & \textbf{Model} & \textbf{MBPP$\uparrow$} & \textbf{MBPP-ET$\uparrow$} & \textbf{HumanEval$\uparrow$} & \textbf{HumanEval-ET$\uparrow$} & \textbf{CodeContests$\uparrow$} & \textbf{Avg$\uparrow$} \\
% \midrule
% \multirow{3}[2]{*}{\textbf{GPT-4o}} & LDB   & 82.4  & 65.4  & 92.1  & 81.7  & 29.3  & 70.2 \\
%       & LPW   & 84.8  & 65.8  & 98.2  & 84.8  & 34.7  & 73.7 \\
%       & \textbf{ThinkCoder(Ours)} & \textbf{90.0} & \textbf{79.2} & 97.0    & \textbf{89.6} & \textbf{40.1} & \textbf{79.2} \\
% \bottomrule
% \end{tabular}}%
%     % }
%     \caption{\textcolor{black}{Pass@1 for LDB, LPW, and ThinkCoder on MBPP, MBPP-ET, HumanEval, HumanEval-ET, and CodeContests benchmarks with GPT-4o as the backbone.}}
%   \label{tab:sota2}%
%   % \vspace{-0.2cm}
% \end{table*}%

\subsection{Model Training and Inference}
Generating high-probability text does not necessarily align well with human preferences, which may lead to biases in various tasks~\cite{becker2024text}. 
Reinforced Self-Training(ReST) mitigates this issue effectively by collecting offline data aligned with human preferences for online training~\cite{gulcehre2023reinforced}. 
We train the base model $\mathcal{M}$ with the success trajectory dataset $D_{M_0} \cup R_{M_0}$:
\vspace{-2mm}
\begin{equation*}
% \resizebox{0.48\textwidth}{!}{
     \begin{aligned}
    \textstyle \mathcal{L}_{MLE}(\theta_{\mathcal{M}}) = -\mathbb{E}_{(x, y) \sim \mathcal{D}_{\mathcal{M}_0} \cup \mathcal{R}_{\mathcal{M}_0}} \log p_{\theta_{\mathcal{M}}}(y \mid x)
    \end{aligned}
% }
\vspace{-2mm}
\end{equation*}
In inference, we compare the performance with baselines at two configurations. The first uses a fine-tuned base model, feeding input $x$ directly to generate text at temperature $t=0$. This setup tests if ThinkCoder data improves output quality.
The second configuration uses the fine-tuned base model as the exploration agent in ThinkCoder. It generates and refines solutions with $k$ explorations and $n$ refinements, using different temperature settings $t$. This aims to assess the framework's ability to improve solutions through successful trajectories and the ThinkCoder framework.
% Comprehensive experimental results are presented in the evaluation section.

\section{Experiments}

\subsection{Dataset}
We utilized the MBPP~\cite{austin2021program} and HumanEval~\cite{chen2021evaluating} datasets to evaluate the effectiveness of our approach. 
To ensure a comprehensive assessment, we tested their enhanced versions—MBPP-ET and HumanEval-ET~\cite{dong2023codescore}—which include 80 and 35 times more test cases than the original datasets, respectively. 
Regarding the test data, the original problem set sizes for MBPP and HumanEval are 257 and 164, while their extended versions contain 378 and 164 problems, respectively. 
For training, we started with 374 MBPP training examples and 90 validation examples, combining them as the initial dataset. 
Using temperature-based and reflection-based generation methods, we expanded this set to over 2,000 examples for ReST training.
\textcolor{black}{We also evaluate our framework on two major benchmarks designed for harder problems: LiveCodeBench and CodeContests. 
LiveCodeBench consists of the 167 most recent samples collected between October 2024 and January 2025, while CodeContests includes 165 problems.}
% To ensure the integrity of the training data, we only retained answers that could be parsed into executable code and corresponding test cases.

\subsection{Baselines}
In terms of LLM base model selection for the application of ThinkCoder, we considered a range of advanced instruction-tuned models, including the general-purpose models LLama2-7B-Chat, Kimi\footnote{https://platform.moonshot.cn}, GPT-4-Turbo(gpt-4-1106-preview) and GPT-4o(gpt-4o-2024-05-13), as well as the code-specific model CodeQwen1.5-7B-Chat. 
We conducted a comprehensive evaluation of ThinkCoder's Pass@1 performance on MBPP and HumanEval across LLMs of different parameter sizes.
For the test-time optimization process, we used the base model LLama2-7B to evaluate performance changes on the MBPP dataset. We compared results obtained before and after fine-tuning, using either the original training data or the success trajectories and incorporating or excluding the ThinkCoder framework.
\textcolor{black}{For the composite model, we compared it with the current SOTA models, MapCoder~\cite{islam2024mapcoder}, AgentCoder~\cite{huang2023agentcoder}, LDB~\cite{zhong2024ldb}, and LPW~\cite{lei2024planning}.}

\begin{table}[htbp]
  \centering
  \resizebox{7.5cm}{!}{
\begin{tabular}{cccc}
\toprule
\textcolor{black}{\textbf{Model}} & \textcolor{black}{\textbf{n}} & \textcolor{black}{\textbf{LiveCodeBench$\uparrow$}} & \textcolor{black}{\textbf{CodeContests$\uparrow$}} \\
\midrule
\multirow{4}[8]{*}{\textcolor{black}{\textbf{GPT-4-Turbo}}} & \textcolor{black}{0}     & \textcolor{black}{33.4}  & \textcolor{black}{16.1} \\
\cmidrule{2-4}      & \textcolor{black}{1}     & \textcolor{black}{43.7(30.8\%)}  & \textcolor{black}{26.1(62.1\%)} \\
\cmidrule{2-4}      & \textcolor{black}{2}     & \textcolor{black}{\textbf{46.1(38.0\%)}}  & \textcolor{black}{27.3(69.6\%)} \\
\cmidrule{2-4}      & \textcolor{black}{5}     & \textcolor{black}{44.9(34.4\%)}  & \textcolor{black}{\textbf{28.5(77.0\%)}}\\
\midrule
\multirow{4}[8]{*}{\textcolor{black}{\textbf{GPT-4o}}} & \textcolor{black}{0}     & \textcolor{black}{22.5}  & \textcolor{black}{21.0} \\
\cmidrule{2-4}      & \textcolor{black}{1}     & \textcolor{black}{35.3(56.9\%)}  & \textcolor{black}{34.1(62.4\%)} \\
\cmidrule{2-4}      & \textcolor{black}{2}     & \textcolor{black}{38.3(70.2\%)}  & \textcolor{black}{38.9(85.2\%)} \\
\cmidrule{2-4}      & \textcolor{black}{5}     & \textcolor{black}{\textbf{41.3(83.6\%)}}  & \textcolor{black}{\textbf{40.1(91.0\%)}} \\
\bottomrule
\end{tabular}%
    }
    \caption{\textcolor{black}{ThinkCoder performance on LiveCodeBench and CodeContests. $n$ represents the self-refinement budget, with the self-exploration budget fixed at $k=5$.}}
  \label{tab:hard}%
\end{table}%

% \begin{table}[htbp]
%   \centering
%   \resizebox{7.5cm}{!}{
% \begin{tabular}{cccc}
% \toprule
% \textbf{Model} & \textbf{n} & \textbf{LiveCodeBench$\uparrow$} & \textbf{CodeContests$\uparrow$} \\
% \midrule
% \multirow{4}[8]{*}{\textbf{GPT-4-Turbo}} & 0     & 33.4  & 16.1 \\
% \cmidrule{2-4}      & 1     & 43.7(30.8\%)  & 26.1(62.1\%) \\
% \cmidrule{2-4}      & 2     & \textbf{46.1(38.0\%)}  & 27.3(69.6\%) \\
% \cmidrule{2-4}      & 5     & 44.9(34.4\%)  & \textbf{28.5(77.0\%)}\\
% \midrule
% \multirow{4}[8]{*}{\textbf{GPT-4o}} & 0     & 22.5  & 21.0 \\
% \cmidrule{2-4}      & 1     & 35.3(56.9\%)  & 34.1(62.4\%) \\
% \cmidrule{2-4}      & 2     & 38.3(70.2\%)  & 38.9(85.2\%) \\
% \cmidrule{2-4}      & 5     & \textbf{41.3(83.6\%)}  & \textbf{40.1(91.0\%)} \\
% \bottomrule
% \end{tabular}%
%     }
%     \caption{\textcolor{black}{ThinkCoder performance on LiveCodeBench and CodeContests. $n$ represents the self-refinement budget, with the self-exploration budget fixed at $k=5$.}}
%   \label{tab:hard}%
% \end{table}%

\begin{figure*}[htb]
  % \begin{wrapfigure}{r}{0.48\textwidth}
  \centering
  \includegraphics[width=1.0\textwidth]{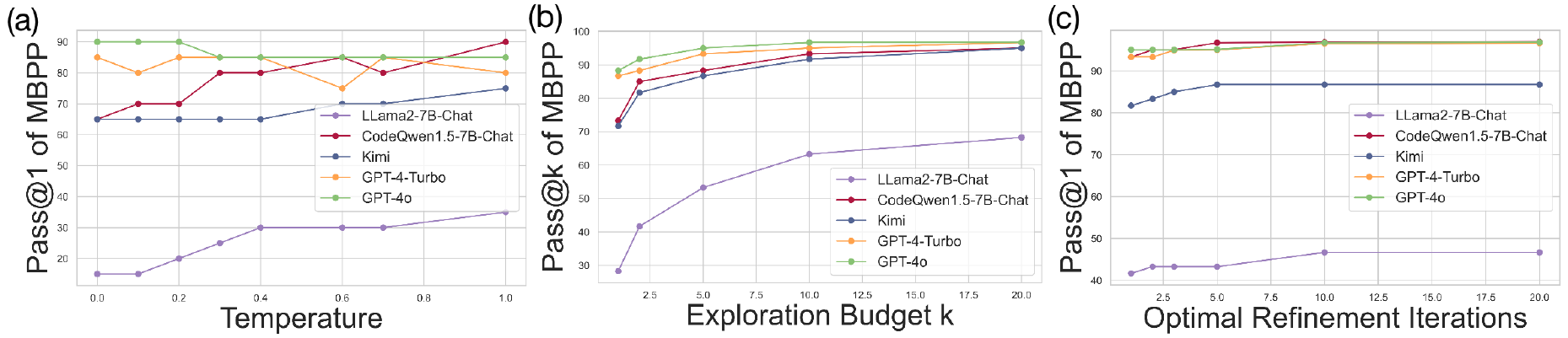}
  \caption{(a) The Pass@1 metric of the baseline models under different temperature $t$.
  (b) The Pass@k metric of the baseline models, where $k$ represents the exploration budget.
  (c) The variation of the Pass@1 metric for each baseline model under the ThinkCoder framework as the optimal refinement budget $n$ increases.}
  \label{fig:parameter}
% \end{wrapfigure}
\end{figure*}

\subsection{Implementation details}
We set different parameter combinations for different experimental settings.  
To validate the improvement of the base LLMs, we set the exploration budget $k = 5$, the temperature received by the exploration agent $t = 0.5$, the number of optimal refinements $n = 5$, and the budget control threshold $\theta = 0.8$.
The exploration agent generates $m=3$ test cases per code.
To compare ThinkCoder with SOTA results, we set the hyperparameters $k$, $t$, $n$, $\theta$, and $m$ to 20, 1.0, 2, 1.0, and 3, respectively. 
In the data collection phase, the exploration agent $\mathcal{M}_0$ is specified as CodeQwen1.5-7B-Chat because it is free and performs well. 
During ReST training, we set the learning rate to $2 \times 10^{-5}$ and train for 2 epochs.
We employ Low-Rank Adaptation (LoRA)~\cite{hu2021lora} to train LLMs.

\section{Evaluation}
\begin{figure}[htb]
  % \begin{wrapfigure}{r}{0.48\textwidth}
  \centering
  \includegraphics[width=0.5\textwidth]{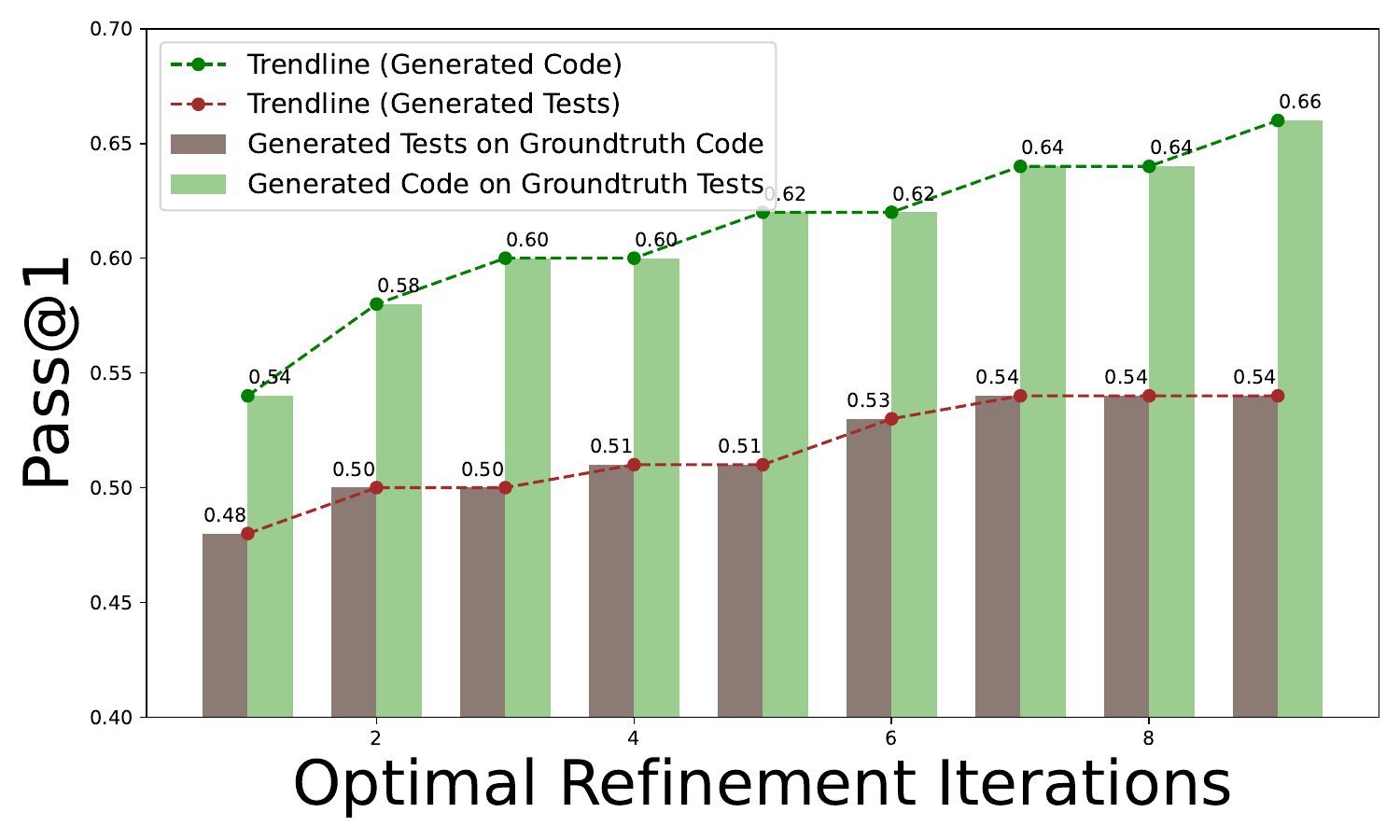}
  \caption{The variation in Pass@1 performance of the exploration code and test cases on the MBPP dataset under ThinkCoder, as evaluated by CodeQwen1.5-7B-Chat as the base LLM.}
  \label{fig:max_acc}
% \end{wrapfigure}
\end{figure}

Table \ref{tab:rpsvr} illustrates the performance changes of various base LLMs on the MBPP and HumanEval datasets as the Optimal Refinement iteration $n$ increases. 
\textcolor{black}{Table \ref{tab:sota1} and Table \ref{tab:sota2} compare the performance of ThinkCoder and SOTA models, using GPT-4-Turbo and GPT-4o as the base LLMs. 
Table \ref{tab:hard} shows great potential on harder benchmarks.}
Table \ref{tab:cost} compares ThinkCoder's computation load with MapCoder at similar settings.
Table \ref{tab:llama} presents the performance changes of LLama2-7B before and after fine-tuning with trajectory data collected by ThinkCoder.
\textcolor{black}{Table \ref{tab:ablation} illustrates the ablation study of self-exploration and self-refinement.}
Figure \ref{fig:parameter} shows the impact of changes in the hyperparameter temperature $t$, exploration budget $k$, and refinement iterations $n$ on the Pass@1 and Pass@k metric.
Figure \ref{fig:max_acc} depicts the trends in the accuracy of generated test cases and codes as the optimal refinement iterations in the ThinkCoder increase.

\subsection{Performance on Code Generation}
Table \ref{tab:rpsvr} shows that ThinkCoder achieves significant improvements in scenarios with fewer standard test cases, with performance on MBPP and HumanEval improving by 4\%-70\% after 5 iterations of refinement. In contrast, gains in more extensive scenarios, like MBPP-ET and HumanEval-ET, are more modest, ranging from 1\%-35\%. 
ThinkCoder’s generated test cases effectively verify code, reducing the need for manual tests, but its performance saturates in complex scenarios, making it better suited for simpler tasks.
Table \ref{tab:sota1} further demonstrates the potential of ThinkCoder. 
When the number of exploration iterations $k$ is extended to 20, ThinkCoder surpasses AgentCoder's performance, which needs 5 refinement rounds, in just 2 iterations.

\begin{table*}[htbp]
  \centering
  \scriptsize
% Table generated by Excel2LaTeX from sheet 'all tables'
\begin{tabular}{ccccccccccccc}
\toprule
\multirow{2}[4]{*}{\textbf{Models}} & \multicolumn{4}{c}{\textbf{HumanEval}} & \multicolumn{4}{c}{\textbf{Mbpp}} & \multicolumn{4}{c}{\textbf{\textcolor{black}{CodeContest}}} \\
\cmidrule{2-13}      & \textbf{n} & \textbf{k} & \textbf{Pass@1}$\uparrow$ & \textbf{Tokens (k)}$\downarrow$ & \textbf{n} & \textbf{k} & \textbf{Pass@1}$\uparrow$ & \textbf{Tokens (k)}$\downarrow$ & \textbf{\textcolor{black}{n}} & \textbf{\textcolor{black}{k}} & \textcolor{black}{\textbf{Pass@1}$\uparrow$} & \textcolor{black}{\textbf{Tokens (k)}$\downarrow$} \\
\midrule
MapCoder & 5     & 5     & 93.9  & 21.3  & 3     & 3     & 83.1  & 5.58  & \textcolor{black}{3}     & \textcolor{black}{3}     & \textcolor{black}{28.5}  & \textcolor{black}{18.3} \\
\midrule
\textbf{ThinkCoder(Ours)} & 20    & 1     & \textbf{93.9}  & \textbf{0.26}  & 5     & 1     & \textbf{89.5}  & \textbf{0.24}  & \textcolor{black}{5}     & \textcolor{black}{5}     & \textcolor{black}{\textbf{28.5}}  & \textcolor{black}{\textbf{2.94}} \\
\bottomrule
\end{tabular}%
    \caption{The token usage for agent responses and the required refinement iterations in ThinkCoder for HumanEval and MBPP, compared to MapCoder (as ThinkCoder and MapCoder share similar configurations), with GPT-4-Turbo as the base LLM. The tiktoken package is used to calculate the response token usage.}
  \label{tab:cost}%
  % \vspace{-0.2cm}
\end{table*}%

\subsection{\textcolor{black}{Ablation Study of Self-exploration and Self-refinement}}
\textcolor{black}{
Figures \ref{fig:parameter}(b) and \ref{fig:parameter}(c) highlight the performance gains from both self-exploration and self-refinement. The gain from self-exploration is greater than that from self-refinement. As illustrated in Table \ref{tab:ablation}, self-refinement significantly enhances efficiency when paired with self-exploration. 
However, increasing the exploration budget excessively (e.g., $k=20$) leads to inefficiency. A moderate increase in the exploration budget accelerates the optimization of the self-refinement process.}

% Table generated by Excel2LaTeX from sheet 'Sheet2'
\begin{table}[htbp]
  \centering
  \resizebox{7.5cm}{!}{
\begin{tabular}{ccccccc}
\toprule
\multirow{2}[4]{*}{\textbf{Models}} &  \multicolumn{6}{c}{\textbf{Optimal Refinement Iterations}} \\
\cmidrule{2-7}      & 0     & 1     & 2     & 3     & 4     & 5 \\
\midrule
\textbf{LLama2-7B} & 14.7  & 22.7  & 24.9  & 27.7  & 32.7  & 36.9 \\
\textbf{LLama2-7B(SFT)}   &   27.8    &   35.3    &    37.5   &  37.6     &  38.0     & 38.1 \\
\textbf{LLama2-7B(ReST)}  & \textbf{32.3}  & \textbf{46.2}  & \textbf{44.8}  & \textbf{46.1}  & \textbf{45.7}  & \textbf{46.1} \\
% \textbf{LLama2-7B(ReST)} & 0.6   & 28.8  & 44.9  & \textbf{45}    & 45.6  & \textbf{45.9}  & \textbf{47.5} \\
\bottomrule
\end{tabular}%
    }
    \caption{Performance comparison of the LLama2-7B model under different settings: baseline without fine-tuning, supervised fine-tuning (SFT) with original train dataset, and Reinforced Self-Training (ReST) with success trajectories from ThinkCoder on the MBPP dataset.}
  \label{tab:llama}%
\end{table}%

\subsection{Performance with ReST Training}
From the results presented in Table \ref{tab:llama}, it is clear that ReST fine-tuning on LLama2-7B using successful trajectories collected by ThinkCoder significantly enhances the model's core capabilities. 
Firstly, LLama2-7B(ReST) demonstrates superior performance compared to both LLama2-7B(SFT) and LLama2-7B at iteration 0, highlighting that ReST training improves the exploration agent's ability to identify better solutions. 
Secondly, with a temperature setting of 0.5, the model achieves optimal performance with only a single round of refinement, underscoring the efficiency of ReST fine-tuning in reducing computational overhead. 
% Finally, when the temperature is increased to 0.6, the model explores a wider range of solutions and achieves a higher Pass@1 score of 47.5 at $n=5$ iterations, showcasing its capacity for more diverse and effective exploration.

\begin{table}[htbp]
  \centering
  \resizebox{7.5cm}{!}{
% Table generated by Excel2LaTeX from sheet 'all tables'
\begin{tabular}{ccccccc}
\toprule
\multicolumn{1}{c}{\multirow{2}[4]{*}{\textcolor{black}{\textbf{\makecell{Self-exploration\\Budget k}}}}} & \multicolumn{6}{c}{\textcolor{black}{\textbf{Self-refinement Budget n}}} \\
\cmidrule{2-7}      & \textcolor{black}{\textbf{1}} & \textcolor{black}{\textbf{2}} & \textcolor{black}{\textbf{3}} & \textcolor{black}{\textbf{5}} & \textcolor{black}{\textbf{10}} & \textcolor{black}{\textbf{20}} \\
\midrule
\textcolor{black}{\textbf{1}} & \textcolor{black}{91.7}  & \textcolor{black}{90.0}    & \textcolor{black}{91.7}  & \textcolor{black}{93.3}  & \textcolor{black}{93.3}  & \textcolor{black}{95.0} \\
\midrule
\textcolor{black}{\textbf{2}} & \textcolor{black}{93.3}  & \textcolor{black}{95.0}    & \textcolor{black}{95.0}    & \textcolor{black}{96.7}  & \textcolor{black}{96.7}  & \textcolor{black}{96.7} \\
\midrule
\textcolor{black}{\textbf{5}} & \textcolor{black}{93.3}  & \textcolor{black}{93.3}  & \textcolor{black}{95.0}    & \textcolor{black}{95.0}    & \textcolor{black}{96.7}  & \textcolor{black}{96.7} \\
\midrule
\textcolor{black}{\textbf{10}} & \textcolor{black}{96.7}  & \textcolor{black}{96.7}  & \textcolor{black}{96.7}  & \textcolor{black}{96.7}  & \textcolor{black}{96.7}  & \textcolor{black}{96.7} \\
\midrule
\textcolor{black}{\textbf{20}} & \textcolor{black}{96.7}  & \textcolor{black}{96.7}  & \textcolor{black}{96.7}  & \textcolor{black}{96.7}  & \textcolor{black}{96.7}  & \textcolor{black}{96.7} \\
\bottomrule
\end{tabular}%
    }
    \caption{\textcolor{black}{The performance trends on the randomly 60 Mbpp samples on GPT-4 with varying compositions of self-exploration and self-refinement budgets.}}
  \label{tab:ablation}%
\end{table}%

% % Table generated by Excel2LaTeX from sheet 'Sheet2'
% \begin{table}[htbp]
%   \centering
%   \resizebox{7.5cm}{!}{
% % Table generated by Excel2LaTeX from sheet 'all tables'
% \begin{tabular}{ccccccc}
% \toprule
% \multicolumn{1}{c}{\multirow{2}[4]{*}{\textbf{\makecell{Self-exploration\\Budget k}}}} & \multicolumn{6}{c}{\textbf{Self-refinement Budget n}} \\
% \cmidrule{2-7}      & \textbf{1} & \textbf{2} & \textbf{3} & \textbf{5} & \textbf{10} & \textbf{20} \\
% \midrule
% \textbf{1} & 91.7  & 90.0    & 91.7  & 93.3  & 93.3  & 95.0 \\
% \midrule
% \textbf{2} & 93.3  & 95.0    & 95.0    & 96.7  & 96.7  & 96.7 \\
% \midrule
% \textbf{5} & 93.3  & 93.3  & 95.0    & 95.0    & 96.7  & 96.7 \\
% \midrule
% \textbf{10} & 96.7  & 96.7  & 96.7  & 96.7  & 96.7  & 96.7 \\
% \midrule
% \textbf{20} & 96.7  & 96.7  & 96.7  & 96.7  & 96.7  & 96.7 \\
% \bottomrule
% \end{tabular}%
%     }
%     \caption{\textcolor{black}{The performance trends on the randomly 60 Mbpp samples on GPT-4 with varying compositions of self-exploration and self-refinement budgets.}}
%   \label{tab:ablation}%
% \end{table}%

\subsection{Evolution of Generated Code and Tests}
After each round of optimal refinement, we evaluated the generated solutions and testing pools. 
Using the ground truth test set, we analyzed the performance changes of the solutions in each round. 
Meanwhile, the ground truth code was used to assess the changes in the testing pools. 
As shown in Figure \ref{fig:max_acc}, the accuracy of both the solutions and testing pools improved simultaneously. 
This demonstrates that ThinkCoder is capable of generating continuously optimized test cases while also refining the quality of the solutions.

\subsection{Hyperparameters Study}
We conducted hyperparameter selection experiments on the MBPP dataset. As shown in Figure \ref{fig:parameter}(a), models exhibit optimal performance at different temperature settings. LLama2-7B-Chat, CodeQwen1.5-7B-Chat, and Kimi improve in Pass@1 as temperature increases, while GPT-4-Turbo and GPT-4o perform better at lower temperatures.
% We conducted hyperparameter experiments on the MBPP dataset. Figure \ref{fig:parameter}(a) shows that models like LLama2-7B-Chat and Kimi improve Pass@1 with higher temperatures, while GPT-4-Turbo and GPT-4o perform better at lower temperatures. 
Figure \ref{fig:parameter}(b) indicates that Pass@k performance of base LLMs improves with increasing exploration budget $k$ but eventually stabilizes. 
Figure \ref{fig:parameter}(c) shows that Pass@1 performance improves with more refinement iterations $n$, but exploration gains are more significant. 
For details on hyperparameters $\theta$ and $m$, see Appendix \ref{sec:hyperparameter} and \ref{sec:param}.
% As shown in Figure \ref{fig:parameter}(c), Pass@1 performance of base LLMs improves with increasing refinement iterations $n$, following a trend similar to the budget curve during exploration. However, exploration consistently achieves greater performance gains, underscoring the importance of optimizing the LLM’s answer selection ability for better code quality.
% For the hyperparameter study of $\theta$ and $m$, please refer to Appendix \ref{sec:hyperparameter} and \ref{sec:param}. 

\subsection{Cost Analysis}
In Table \ref{tab:cost}, we compare the computational overhead between MapCoder and ThinkCoder. \textcolor{black}{Compablue to MapCoder, which uses four LLM-based agents executed serially, ThinkCoder employs only one LLM-based agent and one non-LLM-based CodeVerifier, resulting in lower time overhead. Therefore, we use token overload as a measure of cost consumption.} It reveals that ThinkCoder demonstrates more efficient resource utilization compared to MapCoder with a similar pass@1.

\section{Conclusion}
In this paper, we introduce ThinkCoder, a framework designed to improve the code generation efficiency of LLMs. 
The proposed approach incorporates LLM's exploration with a non-LLM-based CodeVerifier that greatly reduces the token usage. 
To further minimize computational costs during exploration, we capture successful trajectories through ThinkeCoder to fine-tune the LLM. 
This process enhances code generation performance while reducing refinement iterations for optimal solutions.
We evaluate our framework across LLMs of varying sizes and demonstrate its superior performance over current composite models.

\section*{Acknowledgments}
This work is also supported by the Public Computing Cloud, Renmin University of China and by fund for building worldclass universities (disciplines) of Renmin University of China.

\section*{Limitations}
In our current approach to managing the testing pool, we primarily focus on selective expansion but lack effective strategies for cleaning noise from the test set. 
Moreover, test set diversity relies heavily on LLMs' generation capabilities. Our work highlights the importance of tests and the gap between generated and annotated cases. We plan to enhance the testing pool in future work.
Furthermore, while ReST training has demonstrated excellent results in enhancing the code generation capabilities of foundational LLMs, validation in ThinkCoder reveals rapid convergence. 
This indicates that fine-tuning teaches the LLM to find good answers, but the improvement in exploring the overall solution space remains limited.
To address this, we look forward to introducing on-policy training methods in future work, combining data collection with fine-tuning to continuously evolve the model during training. This will generate more diverse answers and enhance the LLM's online exploration capabilities.

\bibliography{acl_latex}

\begin{thebibliography}{36}
\providecommand{\natexlab}[1]{#1}

\bibitem[{Austin et~al.(2021)Austin, Odena, Nye, Bosma, Michalewski, Dohan, Jiang, Cai, Terry, Le et~al.}]{austin2021program}
Jacob Austin, Augustus Odena, Maxwell Nye, Maarten Bosma, Henryk Michalewski, David Dohan, Ellen Jiang, Carrie Cai, Michael Terry, Quoc Le, et~al. 2021.
\newblock Program synthesis with large language models.
\newblock \emph{arXiv preprint arXiv:2108.07732}.

\bibitem[{Backus et~al.(1960)Backus, Bauer, Green, Katz, McCarthy, Perlis, Rutishauser, Samelson, Vauquois, Wegstein et~al.}]{backus1960report}
John~W Backus, Friedrich~L Bauer, Julien Green, Charles Katz, John McCarthy, Alan~J Perlis, Heinz Rutishauser, Klaus Samelson, Bernard Vauquois, Joseph~Henry Wegstein, et~al. 1960.
\newblock Report on the algorithmic language algol 60.
\newblock \emph{Communications of the ACM}, 3(5):299--311.

\bibitem[{Bai et~al.(2023)Bai, Bai, Chu, Cui, Dang, Deng, Fan, Ge, Han, Huang et~al.}]{bai2023qwen}
Jinze Bai, Shuai Bai, Yunfei Chu, Zeyu Cui, Kai Dang, Xiaodong Deng, Yang Fan, Wenbin Ge, Yu~Han, Fei Huang, et~al. 2023.
\newblock Qwen technical report.
\newblock \emph{arXiv preprint arXiv:2309.16609}.

\bibitem[{Becker et~al.(2024)Becker, Wahle, Gipp, and Ruas}]{becker2024text}
Jonas Becker, Jan~Philip Wahle, Bela Gipp, and Terry Ruas. 2024.
\newblock Text generation: A systematic literature review of tasks, evaluation, and challenges.
\newblock \emph{arXiv preprint arXiv:2405.15604}.

\bibitem[{Chaudhary(2023)}]{chaudhary2023code}
Sahil Chaudhary. 2023.
\newblock Code alpaca: An instruction-following llama model for code generation.
\newblock \emph{GitHub repository}.

\bibitem[{Chen et~al.(2023)Chen, Scheurer, Korbak, Campos, Chan, Bowman, Cho, and Perez}]{chen2023improving}
Angelica Chen, J{\'e}r{\'e}my Scheurer, Tomasz Korbak, Jon~Ander Campos, Jun~Shern Chan, Samuel~R Bowman, Kyunghyun Cho, and Ethan Perez. 2023.
\newblock Improving code generation by training with natural language feedback.
\newblock \emph{arXiv preprint arXiv:2303.16749}.

\bibitem[{Chen et~al.(2021)Chen, Tworek, Jun, Yuan, Pinto, Kaplan, Edwards, Burda, Joseph, Brockman et~al.}]{chen2021evaluating}
Mark Chen, Jerry Tworek, Heewoo Jun, Qiming Yuan, Henrique Ponde De~Oliveira Pinto, Jared Kaplan, Harri Edwards, Yuri Burda, Nicholas Joseph, Greg Brockman, et~al. 2021.
\newblock Evaluating large language models trained on code.
\newblock \emph{arXiv preprint arXiv:2107.03374}.

\bibitem[{Dong et~al.(2023)Dong, Ding, Jiang, Li, Li, and Jin}]{dong2023codescore}
Yihong Dong, Jiazheng Ding, Xue Jiang, Ge~Li, Zhuo Li, and Zhi Jin. 2023.
\newblock Codescore: Evaluating code generation by learning code execution.
\newblock \emph{arXiv preprint arXiv:2301.09043}.

\bibitem[{Dong et~al.(2024)Dong, Jiang, Jin, and Li}]{dong2024self}
Yihong Dong, Xue Jiang, Zhi Jin, and Ge~Li. 2024.
\newblock Self-collaboration code generation via chatgpt.
\newblock \emph{ACM Transactions on Software Engineering and Methodology}, 33(7):1--38.

\bibitem[{Gulcehre et~al.(2023)Gulcehre, Paine, Srinivasan, Konyushkova, Weerts, Sharma, Siddhant, Ahern, Wang, Gu et~al.}]{gulcehre2023reinforced}
Caglar Gulcehre, Tom~Le Paine, Srivatsan Srinivasan, Ksenia Konyushkova, Lotte Weerts, Abhishek Sharma, Aditya Siddhant, Alex Ahern, Miaosen Wang, Chenjie Gu, et~al. 2023.
\newblock Reinforced self-training (rest) for language modeling.
\newblock \emph{arXiv preprint arXiv:2308.08998}.

\bibitem[{Guo et~al.(2024)Guo, Cao, Xie, Liu, Li, Chen, and Peng}]{guo2024exploring}
Qi~Guo, Junming Cao, Xiaofei Xie, Shangqing Liu, Xiaohong Li, Bihuan Chen, and Xin Peng. 2024.
\newblock Exploring the potential of chatgpt in automated code refinement: An empirical study.
\newblock In \emph{Proceedings of the 46th IEEE/ACM International Conference on Software Engineering}, pages 1--13.

\bibitem[{Hu et~al.(2021)Hu, Shen, Wallis, Allen-Zhu, Li, Wang, Wang, and Chen}]{hu2021lora}
Edward~J Hu, Yelong Shen, Phillip Wallis, Zeyuan Allen-Zhu, Yuanzhi Li, Shean Wang, Lu~Wang, and Weizhu Chen. 2021.
\newblock Lora: Low-rank adaptation of large language models.
\newblock \emph{arXiv preprint arXiv:2106.09685}.

\bibitem[{Huang et~al.(2023{\natexlab{a}})Huang, Bu, and Cui}]{huang2023codecot}
Dong Huang, Qingwen Bu, and Heming Cui. 2023{\natexlab{a}}.
\newblock Codecot and beyond: Learning to program and test like a developer.
\newblock \emph{arXiv preprint arXiv:2308.08784}.

\bibitem[{Huang et~al.(2023{\natexlab{b}})Huang, Bu, Zhang, Luck, and Cui}]{huang2023agentcoder}
Dong Huang, Qingwen Bu, Jie~M Zhang, Michael Luck, and Heming Cui. 2023{\natexlab{b}}.
\newblock Agentcoder: Multi-agent-based code generation with iterative testing and optimisation.
\newblock \emph{arXiv preprint arXiv:2312.13010}.

\bibitem[{Islam et~al.(2024)Islam, Ali, and Parvez}]{islam2024mapcoder}
Md~Ashraful Islam, Mohammed~Eunus Ali, and Md~Rizwan Parvez. 2024.
\newblock Mapcoder: Multi-agent code generation for competitive problem solving.
\newblock \emph{arXiv preprint arXiv:2405.11403}.

\bibitem[{Jalil et~al.(2023)Jalil, Rafi, LaToza, Moran, and Lam}]{jalil2023chatgpt}
Sajed Jalil, Suzzana Rafi, Thomas~D LaToza, Kevin Moran, and Wing Lam. 2023.
\newblock Chatgpt and software testing education: Promises \& perils.
\newblock In \emph{2023 IEEE international conference on software testing, verification and validation workshops (ICSTW)}, pages 4130--4137. IEEE.

\bibitem[{Jiang et~al.(2024)Jiang, Dong, Wang, Fang, Shang, Li, Jin, and Jiao}]{jiang2024self}
Xue Jiang, Yihong Dong, Lecheng Wang, Zheng Fang, Qiwei Shang, Ge~Li, Zhi Jin, and Wenpin Jiao. 2024.
\newblock Self-planning code generation with large language models.
\newblock \emph{ACM Transactions on Software Engineering and Methodology}, 33(7):1--30.

\bibitem[{Jones et~al.(2024)Jones, Yang, and Eyres}]{jones2024automatic}
Bryan~F Jones, HH~Sthamer~X Yang, and DE~Eyres. 2024.
\newblock The automatic generation of software test data sets using adaptive search techniques.
\newblock \emph{WIT Transactions on Information and Communication Technologies}, 14.

\bibitem[{Lei et~al.(2024)Lei, Chang, Lipovetzky, and Ehinger}]{lei2024planning}
Chao Lei, Yanchuan Chang, Nir Lipovetzky, and Krista~A Ehinger. 2024.
\newblock Planning-driven programming: A large language model programming workflow.
\newblock \emph{arXiv preprint arXiv:2411.14503}.

\bibitem[{Li et~al.(2024{\natexlab{a}})Li, Li, Li, and Jin}]{li2023structured}
Jia Li, Ge~Li, Yongmin Li, and Zhi Jin. 2024{\natexlab{a}}.
\newblock Structured chain-of-thought prompting for code generation.
\newblock \emph{ACM Transactions on Software Engineering and Methodology}.

\bibitem[{Li et~al.(2024{\natexlab{b}})Li, Xia, Du, Dai, Tang, Wang, Yu, and Zhang}]{li2024rethinkmcts}
Qingyao Li, Wei Xia, Kounianhua Du, Xinyi Dai, Ruiming Tang, Yasheng Wang, Yong Yu, and Weinan Zhang. 2024{\natexlab{b}}.
\newblock Rethinkmcts: Refining erroneous thoughts in monte carlo tree search for code generation.
\newblock \emph{arXiv preprint arXiv:2409.09584}.

\bibitem[{Li et~al.(2023)Li, Zong, Wang, Tian, Wang, Cheung, and Kramer}]{li2023finding}
Tsz-On Li, Wenxi Zong, Yibo Wang, Haoye Tian, Ying Wang, Shing-Chi Cheung, and Jeff Kramer. 2023.
\newblock Finding failure-inducing test cases with chatgpt.
\newblock \emph{arXiv preprint arXiv:2304.11686}.

\bibitem[{Liu et~al.(2024{\natexlab{a}})Liu, Le-Cong, Widyasari, Tantithamthavorn, Li, Le, and Lo}]{liu2024refining}
Yue Liu, Thanh Le-Cong, Ratnadira Widyasari, Chakkrit Tantithamthavorn, Li~Li, Xuan-Bach~D Le, and David Lo. 2024{\natexlab{a}}.
\newblock Refining chatgpt-generated code: Characterizing and mitigating code quality issues.
\newblock \emph{ACM Transactions on Software Engineering and Methodology}, 33(5):1--26.

\bibitem[{Liu et~al.(2024{\natexlab{b}})Liu, Chen, Zhang, Gao, Zhang, and Yan}]{liu2024skepticism}
Yuhan Liu, Xiuying Chen, Xiaoqing Zhang, Xing Gao, Ji~Zhang, and Rui Yan. 2024{\natexlab{b}}.
\newblock From skepticism to acceptance: simulating the attitude dynamics toward fake news.
\newblock In \emph{Proceedings of the Thirty-Third International Joint Conference on Artificial Intelligence}, pages 7886--7894.

\bibitem[{Liu et~al.(2024{\natexlab{c}})Liu, Song, Zhang, Chen, and Yan}]{liu2024tiny}
Yuhan Liu, Zirui Song, Xiaoqing Zhang, Xiuying Chen, and Rui Yan. 2024{\natexlab{c}}.
\newblock From a tiny slip to a giant leap: An llm-based simulation for fake news evolution.
\newblock \emph{arXiv preprint arXiv:2410.19064}.

\bibitem[{Muennighoff et~al.(2023)Muennighoff, Liu, Zebaze, Zheng, Hui, Zhuo, Singh, Tang, Von~Werra, and Longpre}]{muennighoff2023octopack}
Niklas Muennighoff, Qian Liu, Armel Zebaze, Qinkai Zheng, Binyuan Hui, Terry~Yue Zhuo, Swayam Singh, Xiangru Tang, Leandro Von~Werra, and Shayne Longpre. 2023.
\newblock Octopack: Instruction tuning code large language models.
\newblock \emph{arXiv preprint arXiv:2308.07124}.

\bibitem[{Qin et~al.(2023)Qin, Liang, Ye, Zhu, Yan, Lu, Lin, Cong, Tang, Qian et~al.}]{qin2023toolllm}
Yujia Qin, Shihao Liang, Yining Ye, Kunlun Zhu, Lan Yan, Yaxi Lu, Yankai Lin, Xin Cong, Xiangru Tang, Bill Qian, et~al. 2023.
\newblock Toolllm: Facilitating large language models to master 16000+ real-world apis.
\newblock \emph{arXiv preprint arXiv:2307.16789}.

\bibitem[{Ridnik et~al.(2024)Ridnik, Kredo, and Friedman}]{ridnik2024code}
Tal Ridnik, Dedy Kredo, and Itamar Friedman. 2024.
\newblock Code generation with alphacodium: From prompt engineering to flow engineering.
\newblock \emph{arXiv preprint arXiv:2401.08500}.

\bibitem[{Roziere et~al.(2023)Roziere, Gehring, Gloeckle, Sootla, Gat, Tan, Adi, Liu, Sauvestre, Remez et~al.}]{roziere2023code}
Baptiste Roziere, Jonas Gehring, Fabian Gloeckle, Sten Sootla, Itai Gat, Xiaoqing~Ellen Tan, Yossi Adi, Jingyu Liu, Romain Sauvestre, Tal Remez, et~al. 2023.
\newblock Code llama: Open foundation models for code.
\newblock \emph{arXiv preprint arXiv:2308.12950}.

\bibitem[{Snell et~al.(2024)Snell, Lee, Xu, and Kumar}]{snell2024scaling}
Charlie Snell, Jaehoon Lee, Kelvin Xu, and Aviral Kumar. 2024.
\newblock Scaling llm test-time compute optimally can be more effective than scaling model parameters.
\newblock \emph{arXiv preprint arXiv:2408.03314}.

\bibitem[{Wang et~al.(2024)Wang, Huang, Chen, Liu, Wang, and Wang}]{wang2024software}
Junjie Wang, Yuchao Huang, Chunyang Chen, Zhe Liu, Song Wang, and Qing Wang. 2024.
\newblock Software testing with large language models: Survey, landscape, and vision.
\newblock \emph{IEEE Transactions on Software Engineering}.

\bibitem[{Zhang et~al.(2024)Zhang, Cheng, Wu, and Hu}]{zhang2024pair}
Huan Zhang, Wei Cheng, Yuhan Wu, and Wei Hu. 2024.
\newblock A pair programming framework for code generation via multi-plan exploration and feedback-driven refinement.
\newblock In \emph{Proceedings of the 39th IEEE/ACM International Conference on Automated Software Engineering}, pages 1319--1331.

\bibitem[{Zhang et~al.(2025)Zhang, Liu, Liu, Luan, Yan et~al.}]{zhang2025weaving}
Juntian Zhang, Yuhan Liu, Wei Liu, Jian Luan, Rui Yan, et~al. 2025.
\newblock Weaving context across images: Improving vision-language models through focus-centric visual chains.
\newblock \emph{arXiv preprint arXiv:2504.20199}.

\bibitem[{Zheng et~al.(2024)Zheng, Zhang, Shen, Liu, Lin, Fu, Chen, and Yue}]{zheng2024opencodeinterpreter}
Tianyu Zheng, Ge~Zhang, Tianhao Shen, Xueling Liu, Bill~Yuchen Lin, Jie Fu, Wenhu Chen, and Xiang Yue. 2024.
\newblock Opencodeinterpreter: Integrating code generation with execution and refinement.
\newblock \emph{arXiv preprint arXiv:2402.14658}.

\bibitem[{Zheng et~al.(2023)Zheng, Ning, Wang, Zhang, Zheng, Ye, and Chen}]{zheng2023survey}
Zibin Zheng, Kaiwen Ning, Yanlin Wang, Jingwen Zhang, Dewu Zheng, Mingxi Ye, and Jiachi Chen. 2023.
\newblock A survey of large language models for code: Evolution, benchmarking, and future trends.
\newblock \emph{arXiv preprint arXiv:2311.10372}.

\bibitem[{Zhong et~al.(2024)Zhong, Wang, and Shang}]{zhong2024ldb}
Li~Zhong, Zilong Wang, and Jingbo Shang. 2024.
\newblock Ldb: A large language model debugger via verifying runtime execution step-by-step.
\newblock \emph{arXiv preprint arXiv:2402.16906}.

\end{thebibliography}

\appendix
\newpage
\section{Appendix}

\subsection{Algorithm of ThinkCoder}
\label{sec:algorithm}
Algorithm \ref{alg:thinkcoder} shows the pseudo-code of ThinkCoder.

\subsection{Exploration Budget Control}
\label{sec:hyperparameter}

\begin{figure}[htb]
  % \begin{wrapfigure}{r}{0.48\textwidth}
  \centering
  \includegraphics[width=0.5\textwidth]{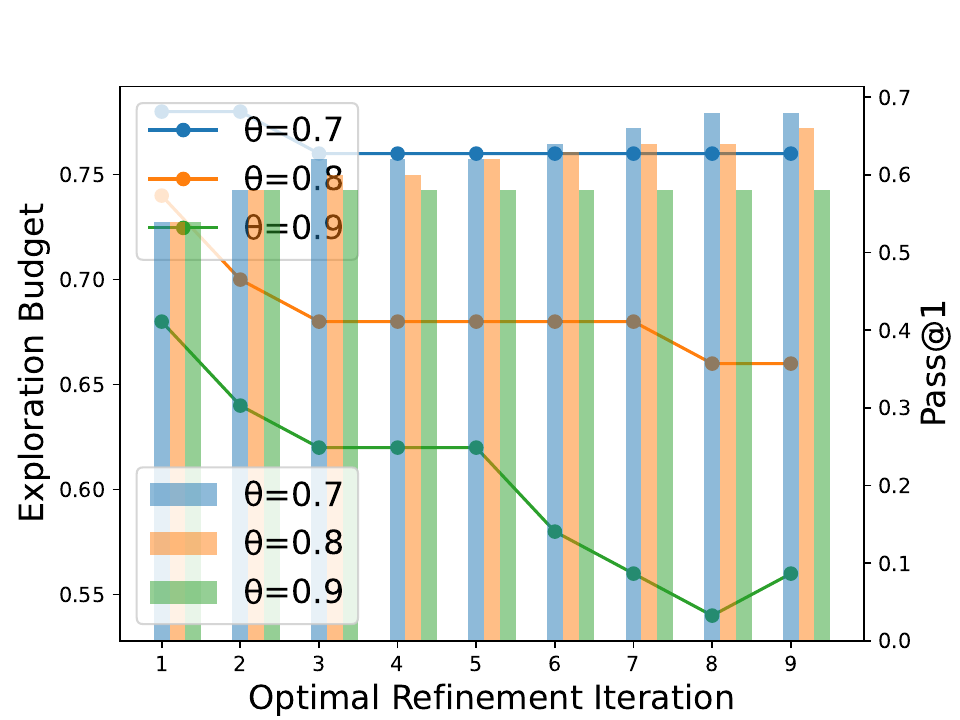}
  \caption{The relationship between the computational cost and performance of ThinkCoder at different budget control thresholds, where `Pass@1' indicates the performance on the MBPP dataset, and `$\theta$' refers to the budget control threshold that allows the task to execute the next exploration process. `Exploration Budget' represents the ratio of total requests during each iteration.}
  \label{fig:budget}
% \end{wrapfigure}
\end{figure}

Figure \ref{fig:budget} outlines the number of tasks for the exploration agent per iteration and the overall Pass@1 metric under different budget control thresholds. 
By increasing the exploration budget control threshold $\theta$, the number of tasks requiring correction in each iteration is significantly reduced, leading to a notable decrease in ThinkCoder's runtime. 
However, this approach risks leaving some incorrectly validated tasks uncorrected, which could negatively impact performance. 
In our experiments, we set $\theta = 0.8$, as this value not only progressively improves Pass@1 but also reduces the number of tasks per iteration by nearly half.

\begin{algorithm}[tb]
\small
\caption{ThinkCoder}
\label{alg:thinkcoder}
\begin{algorithmic}[1]

\State Initialize the testing pool $\text{TP=[]}$, set the globally max pass rate $g_p=0$ and best solution $g_s=\text{None}$
\State Set the hyperparameters $t=0.5, k=5, n=5, m=3, \theta=0.8$
\State Set the instruction $\text{prompt}$ for code generation task and instruction for task $p$ as $I=[p, \text{prompt}]$
\For {each $iter \in n$}
    \State Initialize the locally max pass rate $l_p=0$ and best solution ${l_s}={\text{None}}$ for iteration $iter$
    \State Randomly select $m$ successful test cases from $\text{TP}$ as $\text{tests}$
    \State Select a new type of error feedback for test $f_t$ from $\text{TP}$ as $f$
    \If{$f_t$ and $f$ is not None}
        \State Update the instruction $I=[p, g_s, f_t, f, \text{tests}, \text{prompt}]$
    \EndIf
    \State Get code solutions $\text{G}'$ and test cases $\text{TP}'$ from exploration agent for the instruction $I$ within $k$ iterations
    \State Deduplicate testing pool $\text{TP'}$ by AST
    \For {each $i \in |\text{G}'|$}
        \State Get the pass rate $r_{g_i}$ for solution $g_i$ from CodeVerifier with the input $\text{TP}'$
        \If{$r_{g_i}>l_p$}
            \State Set $l_p=r_{g_i}, l_s={g_i}$
        \EndIf
    \EndFor
    \If{$l_p>g_p$}
        \State Set $g_p=l_p, g_s=l_s$
        \State Add $\text{TP'}$ to $\text{TP}$
    \EndIf
    \If{$g_p > \theta$}
        \State return $g_s$ as the best solution
        \State Terminate optimal refinement iteration
    \EndIf
\EndFor
\end{algorithmic}
\end{algorithm}

\subsection{Exploration for Tests}
\label{sec:param}

\begin{table}[htbp]
  \centering
  \small
  % \resizebox{7.5cm}{!}{
% Table generated by Excel2LaTeX from sheet 'all tables'
% Table generated by Excel2LaTeX from sheet 'mbpp'
\begin{tabular}{lrrrr}
\toprule
\multirow{2}[4]{*}{\textbf{Metrics}} & \multicolumn{4}{c}{\textbf{Tests m/Exploration}} \\
\cmidrule{2-5}      & \textbf{1} & \textbf{3} & \textbf{5} & \textbf{10} \\
\midrule
\textbf{Pass@1} & 80.0    & 81.7  & 81.7  & 86.7\\
\midrule
\textbf{Overload(s)} & 28.7  & 86.0    & 112.3 & 364.5\\
\bottomrule
\end{tabular}%
    % }
    \caption{The variation in the Pass@1 metric and the CodeVerifier's overload is influenced by changes in the number of test cases $m$ during each exploration.}
  \label{tab:m}%
  % \vspace{-0.2cm}
\end{table}%

We randomly selected 60 tasks from the MBPP test data to observe the performance improvement and code execution overhead as the number of test cases $m$ increases during each exploration. 
Table \ref{tab:m} illustrates this trend. 
The base model we used is Kimi. We observed that increasing $m$ leads to improved Pass@1 performance and results in longer execution times due to the additional effort required for self-verification. 
We set $m$ for all experiments to balance time efficiency and performance.

\subsection{Prompts for Exploration Agent}
\label{sec:prompts}
\begin{figure*}[htb]
  % \begin{wrapfigure}{r}{0.48\textwidth}
  \centering
  \includegraphics[width=0.9\textwidth]{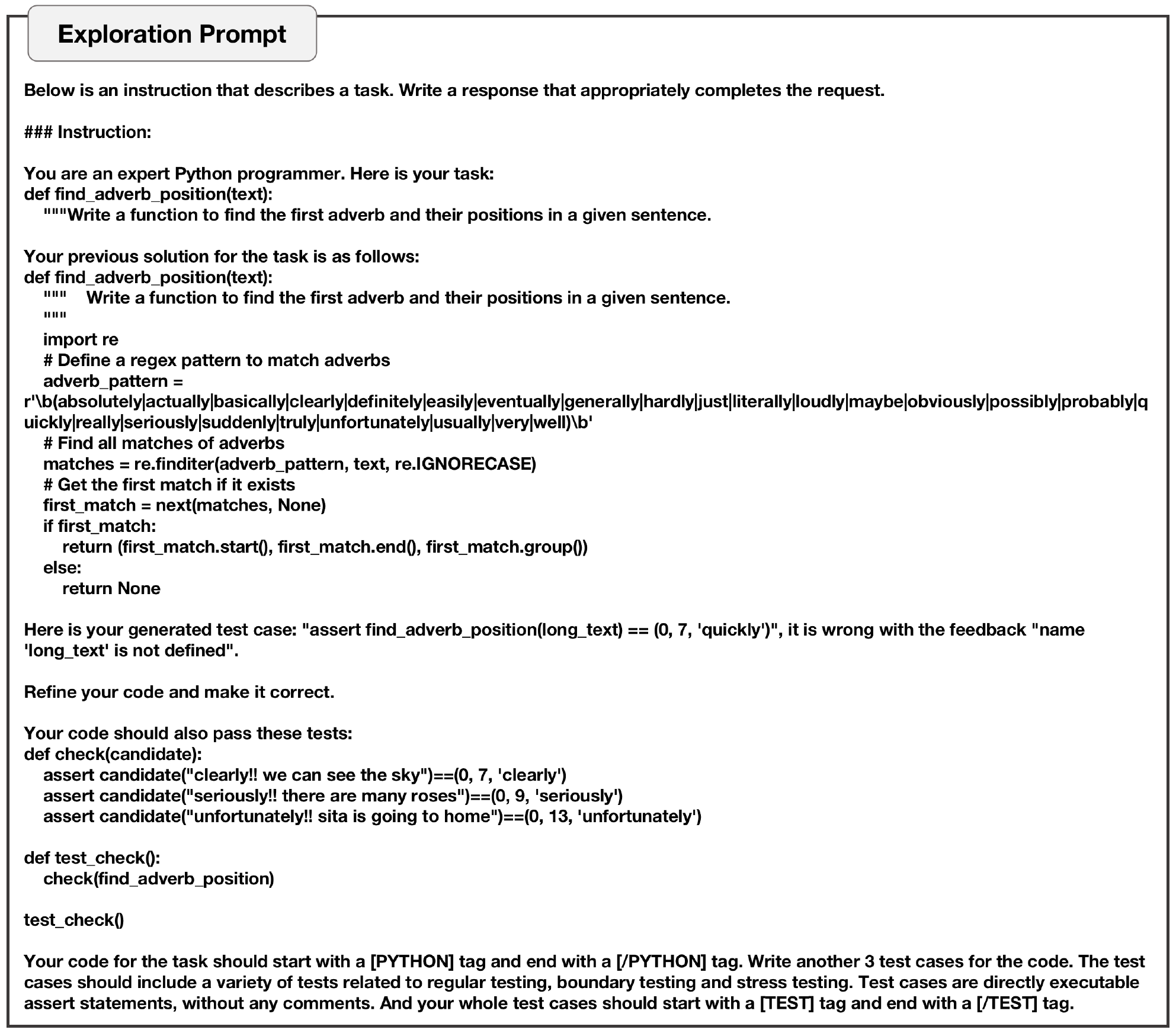}
  \caption{ThinkCoder exploration agent prompt example.}
  \label{fig:explore_prompt}
% \end{wrapfigure}
\end{figure*}

Figure \ref{fig:explore_prompt} illustrates the input received by the exploration agent, including the problem, the best solution, the test cases from the ground truth, the generated test cases from testing pooling, and feedback on test failures.

\subsection{Inputs for CodeVerifier}

\begin{figure*}[htb]
  % \begin{wrapfigure}{r}{0.48\textwidth}
  \centering
  \includegraphics[width=0.9\textwidth]{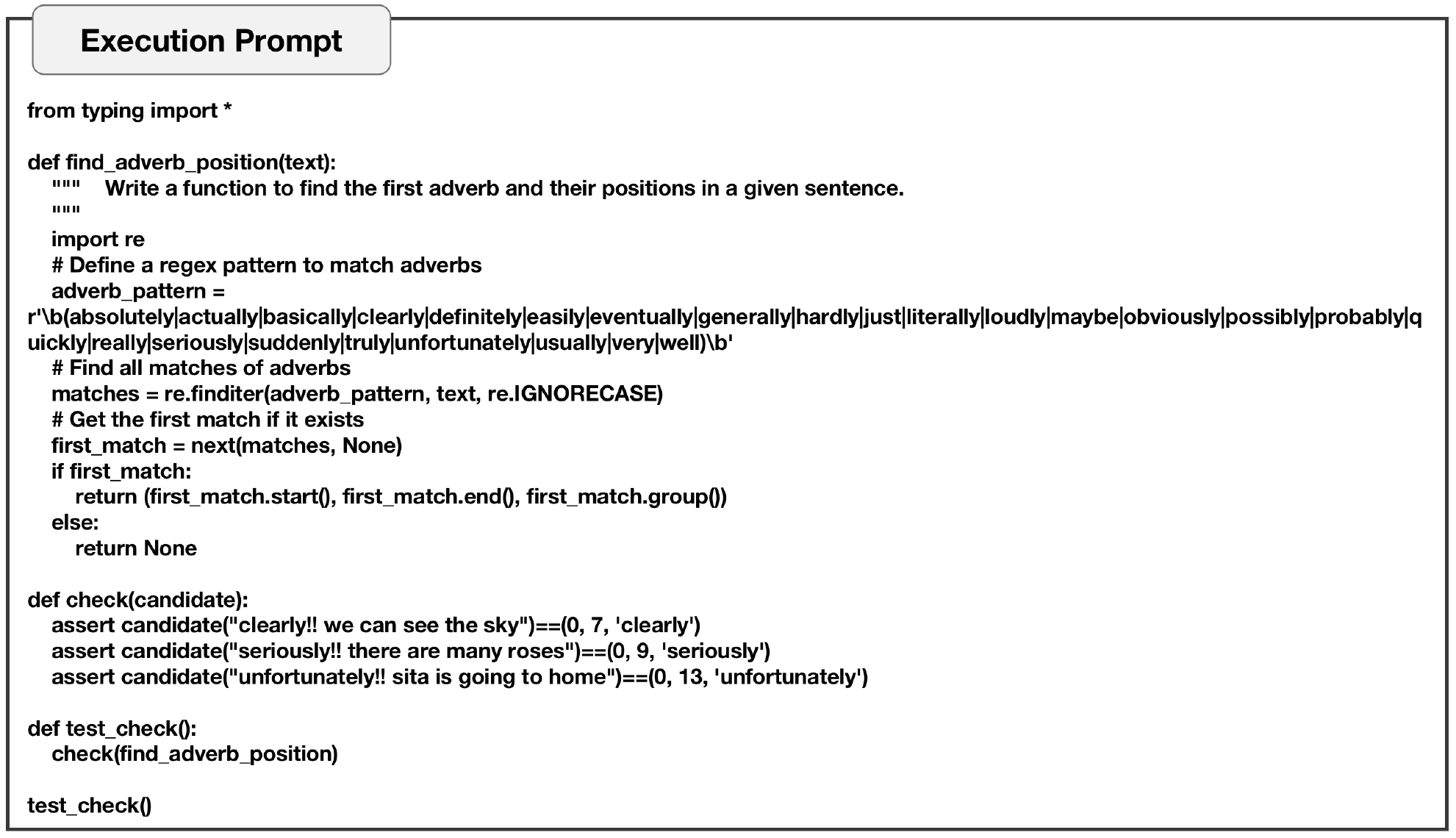}
  \caption{ThinkCoder CodeVerifier input example.}
  \label{fig:execution_prompt}
% \end{wrapfigure}
\end{figure*}

Figure \ref{fig:execution_prompt} shows the input of the executable Python environment, including the code and all test cases from the testing pool.

\subsection{Example of Thorough Exploration}

\begin{figure*}[htb]
  % \begin{wrapfigure}{r}{0.48\textwidth}
  \centering
  \includegraphics[width=0.9\textwidth]{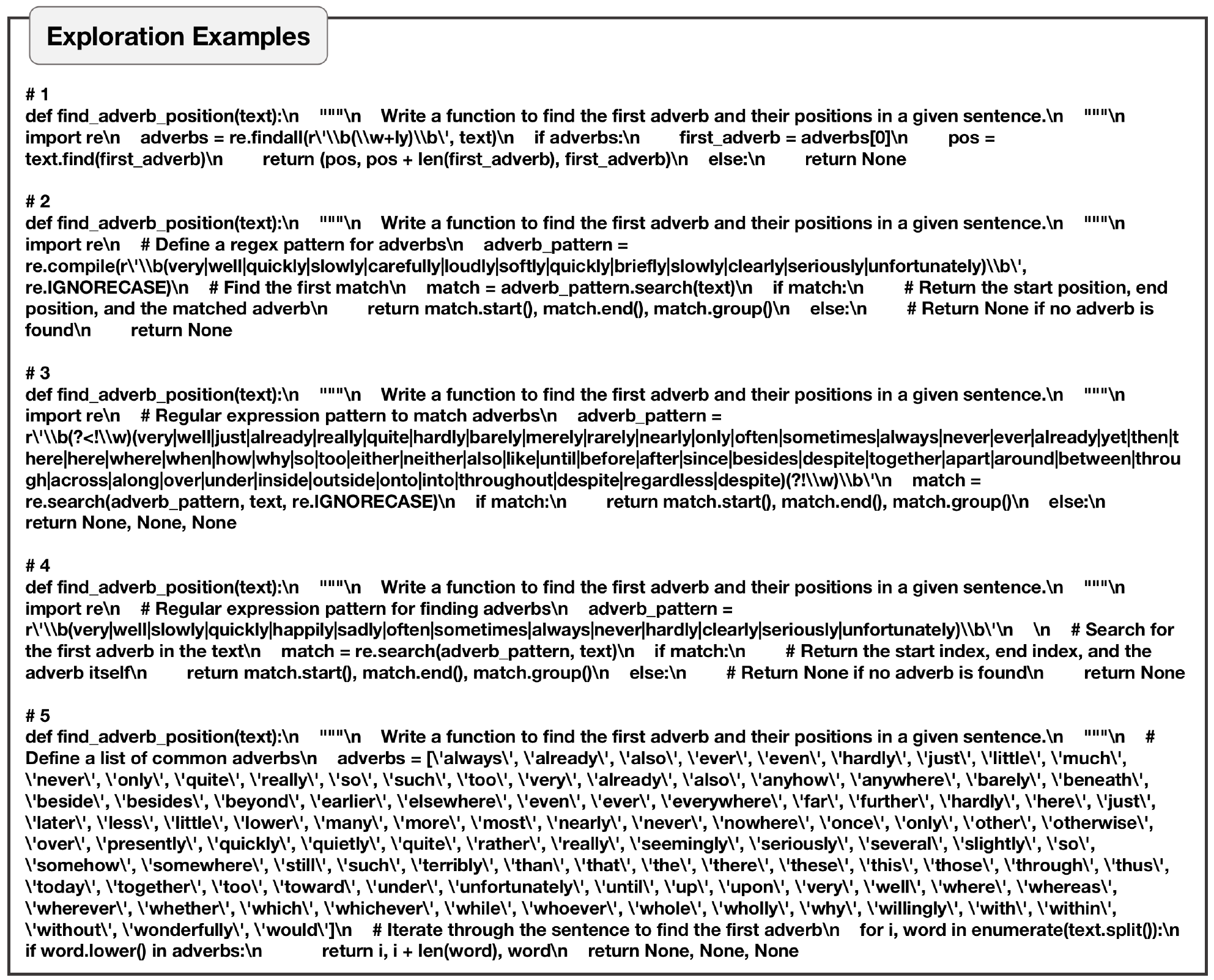}
  \caption{ThinkCoder exploration examples.}
  \label{fig:exploration_example}
% \end{wrapfigure}
\end{figure*}

Figure \ref{fig:exploration_example} illustrates multiple solutions explored by the exploration agent.

\subsection{Example of Optimal Refinement}

\begin{figure*}[htb]
  % \begin{wrapfigure}{r}{0.48\textwidth}
  \centering
  \includegraphics[width=0.9\textwidth]{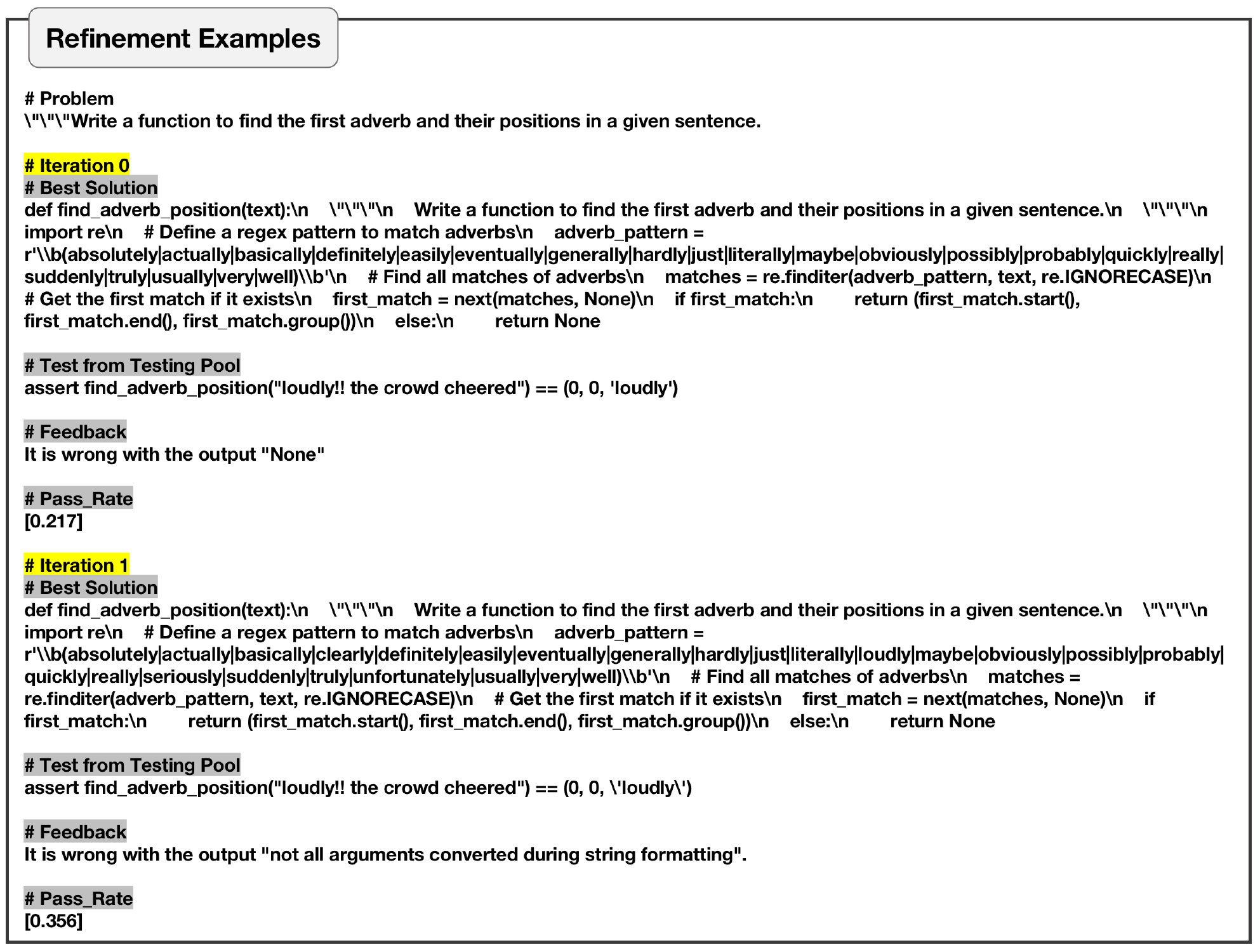}
  \caption{ThinkCoder refinement examples.}
  \label{fig:refinement_example}
% \end{wrapfigure}
\end{figure*}

To better understand the process of optimal refinement, Figure \ref{fig:refinement_example} illustrates a single refinement iteration. 
Before refinement, we leveraged the best solution output from the previous iteration, along with failed tests and unmet feedback sampled from the testing pool. 
Using this reflection information, the solution was updated and revalidated on the test cases in the testing pool. 
The success rate improved from 0.217 to 0.356, leading us to select the current solution as the best solution.

\subsection{Example of Testing Pool}

\begin{figure*}[htb]
  % \begin{wrapfigure}{r}{0.48\textwidth}
  \centering
  \includegraphics[width=0.9\textwidth]{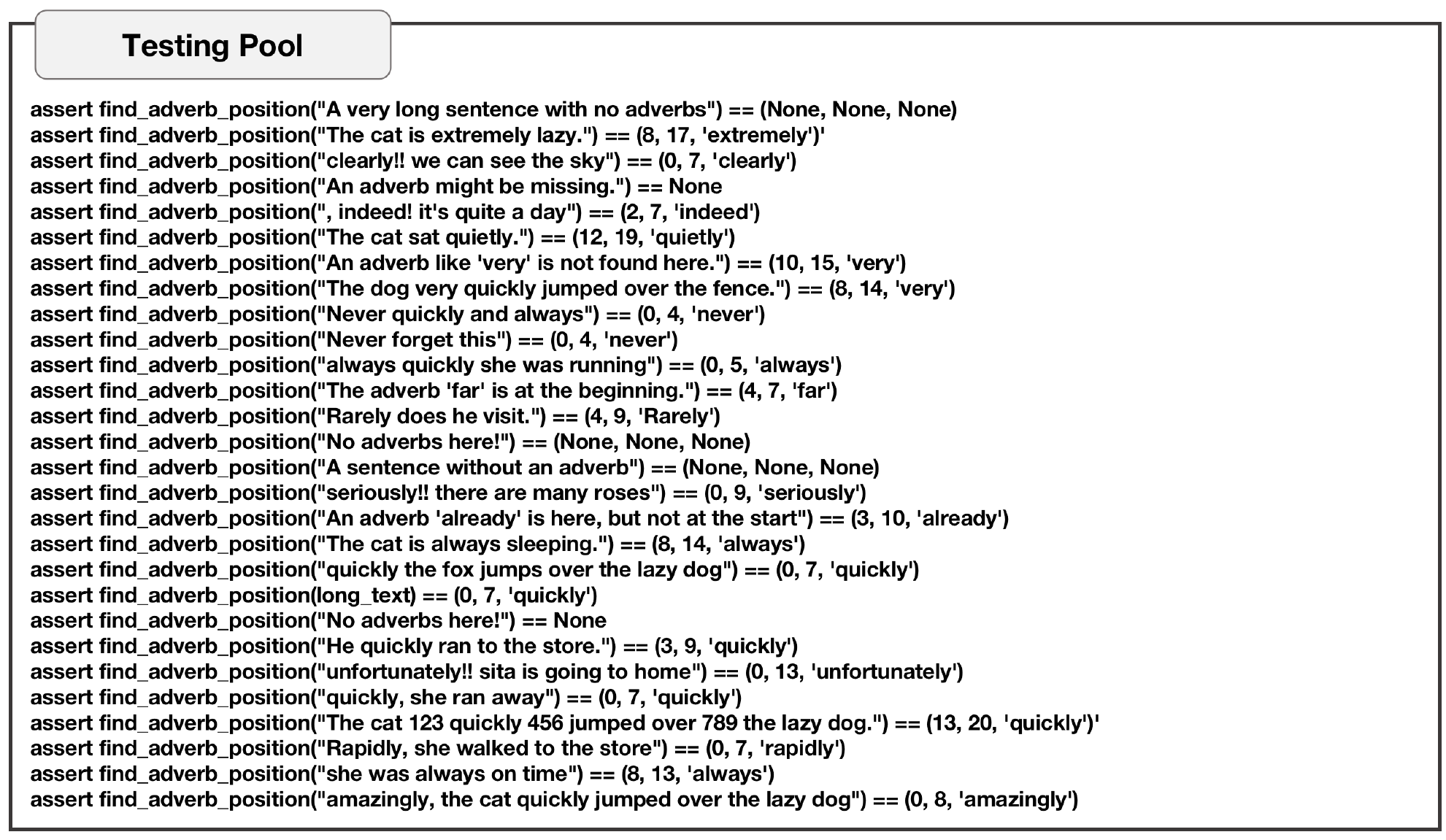}
  \caption{ThinkCoder testing pool examples.}
  \label{fig:testing_pool}
% \end{wrapfigure}
\end{figure*}

Figure \ref{fig:testing_pool} illustrates the test cases in the testing pool generated through multiple explorations by ThinkCoder.

% This is an appendix.

\end{document}